\documentclass[bibyear]{aa}  
\usepackage{graphicx}
\usepackage{txfonts}
\usepackage[outercaption,rightcaption]{sidecap}
\sidecaptionvpos{figure}{c}
\usepackage{array}
\newcolumntype{L}[1]{>{\raggedright\let\newline\\\arraybackslash\hspace{0pt}}m{#1}}
\newcolumntype{C}[1]{>{\centering\let\newline\\\arraybackslash\hspace{0pt}}m{#1}}
\newcolumntype{R}[1]{>{\raggedleft\let\newline\\\arraybackslash\hspace{0pt}}m{#1}}
\usepackage{natbib,twoopt}
\AtBeginDocument{} % Required by "hyperref" when Using \label inside an align environment. 
\usepackage[breaklinks=true,colorlinks=true,allcolors=blue]{hyperref} %% to avoid \citeads line fills
\makeatletter
\newcommandtwoopt{\citeads}[3][][]{\href{http://adsabs.harvard.edu/abs/#3}%
{\def\hyper@linkstart##1##2{}%
\let\hyper@linkend\@empty\citealp[#1][#2]{#3}}}
\newcommandtwoopt{\citepads}[3][][]{\href{http://adsabs.harvard.edu/abs/#3}%
{\def\hyper@linkstart##1##2{}%
\let\hyper@linkend\@empty\citep[#1][#2]{#3}}}
\newcommandtwoopt{\citetads}[3][][]{\href{http://adsabs.harvard.edu/abs/#3}%
{\def\hyper@linkstart##1##2{}%
\let\hyper@linkend\@empty\citet[#1][#2]{#3}}}
\newcommandtwoopt{\citeyearads}[3][][]%
{\href{http://adsabs.harvard.edu/abs/#3}
{\def\hyper@linkstart##1##2{}%
\let\hyper@linkend\@empty\citeyear[#1][#2]{#3}}}
\makeatother
\bibpunct{(}{)}{;}{a}{}{,} %% natbib format for A&A and ApJ
\usepackage{xcolor}

\newcommand{\ha}{H$\alpha$}
\newcommand{\be}{\begin{equation}}
\newcommand{\ee}{\end{equation}}
\newcommand{\bea}{\begin{eqnarray}}
\newcommand{\eea}{\end{eqnarray}}

\graphicspath{ {./}{../images/}}

\begin{document} 

\authorrunning{C. E. Alissandrakis etal}
\title{First detection of metric emission from a solar surge
}
\author{C. E. Alissandrakis\inst{1}, S. Patsourakos\inst{1}, A. Nindos\inst{1}, C. Bouratzis\inst{2}
\and
 A. Hillaris\inst{2}
}
\offprints{C. Alissandrakis}
\institute{Department of Physics, University of Ioannina, GR-45110 Ioannina, 
Greece\\
\email{calissan@uoi.gr}
\and{Section of Astrophysics, Astronomy \& Mechanics, Department of Physics, University of Athens, Panepistimiopolis 157 84, Zografos, Greece}
}

\date{Received ...; accepted ...}

% \abstract{}{}{}{}{} 
% 5 {} token are mandatory
 
\abstract
{We report the first detection of metric radio emission from a surge, observed with the Nan\c cay Radioheliograph (NRH), STEREO and other instruments. The emission was observed during the late phase of the M9 complex event SOL2010-02-012T11:25:00, described in a previous publication and was associated with a secondary energy release, also observed in STEREO 304\,\AA\ images: there was no detectable soft X-ray emission. Triangulation of the STEREO images allowed the identification of the surge with NRH sources near the central meridian. The radio emission of the surge occurred in two phases and consisted of two sources, one located near the base of the surge, apparently at or near the site of energy release, and another in the upper part of the surge; these were best visible in the frequency range of 445.0 to about 300\,MHz, whereas a spectral component of different nature was observed at lower frequencies. Sub-second time variations were detected in both sources during both phases, with 0.2-0.3\,s a delay of the upper source with respect to the lower, suggesting superluminal velocities. This effect can be explained if the emission of the upper source was due to scattering of radiation from the source at the base of the surge. In addition, the radio emission showed signs of pulsations and spikes. We discuss possible emission mechanisms for the slow time variability component of the lower radio source. Gyrosynchrotron emission reproduced fairly well the characteristics of the observed total intensity spectrum at the start of the second phase of the event, but failed to reproduce the high degree of the observed circular polarization as well the spectra at other instances. On the other hand, type IV-like plasma emission from the fundamental could explain the high polarization and the fine structure in the dynamic spectrum; moreover, it gives projected radio source positions on the plane of the sky, as seen from STEREO-A, near the base of the surge. Taking everything into consideration, we suggest type IV-like plasma emission with a low intensity gyrosynchrotron component as the most plausible mechanism.}


   \maketitle
%
%________________________________________________________________

\section{Introduction}
The solar electromagnetic radiation extends from $\gamma$-rays to km wavelengths. Each spectral range provides specific information on solar phenomena, whereas combined observations in many spectral ranges are required to obtain a complete picture of physical processes operating in the solar atmosphere. The solar radio emission in particular, which originates from the low chromosphere at mm-wavelengths up to the outer corona and the solar wind at km wavelengths, covers the same altitude range as emissions in the optical, EUV and X-ray ranges and in situ measurements, hence it is a necessary complement of data obtained in these domains (see, e.g. the recent review by \citeads{2020FrASS...7...74A}). Thus practically all phenomena observed in radio wavelengths have their counterpart in other regions of the electromagnetic spectrum.

Metric solar bursts are an exception to the above statement. Although we have a basic understanding of their origin, some ot their counterparts in other spectral ranges have not been identified. For example, we know that type III bursts are due to beams of accelerated electrons but such beams are not detectable in optical or EUV wavelengths, although they have been observed by particle detectors in the interplanetary space. Similarly, we know that type II bursts are due to shock waves, but we are not sure which observable manifestation of the shocks corresponds to the radio emission. The reason for this lies apparently in the fact that both type IIs and type IIIs are due to coherent plasma emission, which has no observable signature outside the radio range. 

The only definitive association between metric burst and optical/EUV emission is that of the so called ``radio CMEs'', emitting in the radio range through the incoherent gyrosynchrotron mechanism (see, for example~\citeads{2001ApJ...558L..65B}; \citeads{2007ApJ...660..874M}; \citeads{2012ApJ...750..147D}; \citeads{2013ApJ...766..130T}; \citeads{2014ApJ...782...43B}; \citeads{2017A&A...608A.137C}; \citeads{2020ApJ...893...28M}; \citeads{2021ApJ...906..132C}; see also Sect. 5 in \citeads{2021FrASS...7...77A}). This gyrosynchrotron emission originates at the core of the CME, as one of the components of the radio signature which appears on dynamic spectra in the form of a type IV continuum \citepads[see~][~for example]{2021ApJ...906..132C}; additional components of this kind of type IV burst due to plasma radiation (\citeads{2019ApJ...870...30V}; \citeads{2019A&A...623A..63M}) are also possible.

\begin{figure}[]
\includegraphics[width=.92\hsize]{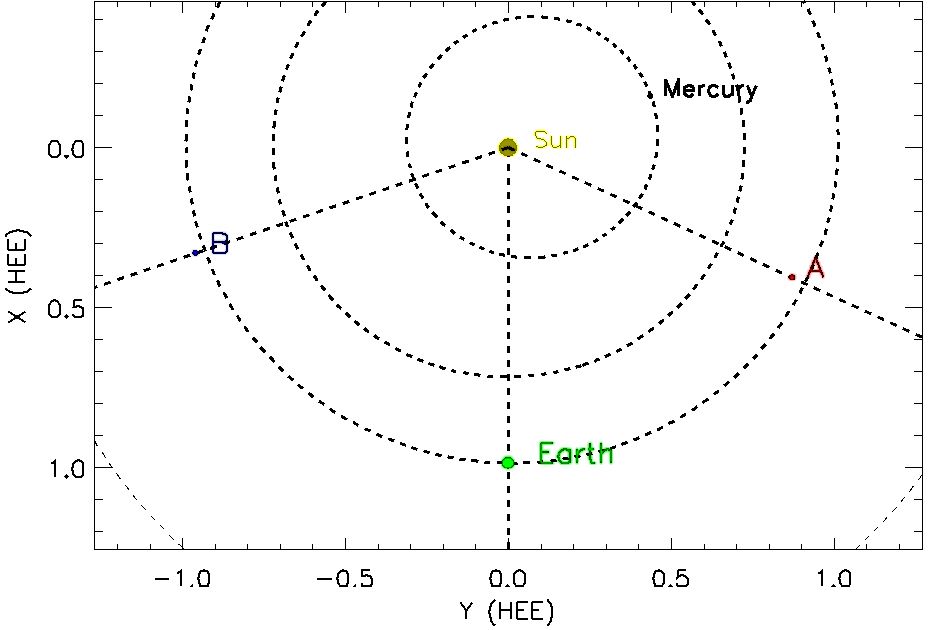}
\caption{{ Positions of the STEREO spacecrafts on February 12, 2010.}}
\label{STEREOpos}
\end{figure}

There have also been reports of soft X-ray (SXR) jets associated with Type III emission, apparently from accelerated electrons that traveled along the path outlined by the jet (\citeads{1995ApJ...447L.135K}; \citeads{1996A&A...306..299R}). Moreover, \citetads{1996ApJ...472..874R} reported the occurrence of a metric continuum source that was associated with both the partial rupture of a coronal loop top and soft X-ray plasma ejecta, whereas \citetads{2001ApJ...559..443K} reported the detection of metric continuum sources associated with soft X-ray plasmoid ejecta. { Earlier works (\citeads{1973SoPh...29..163A}, \citeads{1985SoPh...95..331M}, \citeads{1986SoPh..103..235C}) reported the association of Type IIIs with \ha\ ejecta, surges in particular.} However, no direct emission from the SXR { or optical} structures themselves was reported  by the above authors, with the possible exception of \citetads{1996ApJ...472..874R}.

We note that ``jet'' is a generic term that may describe material ejections of various nature, including surges. The latter, classically, are ejections of chromospheric material, best observed in \ha, usually seen in absorption on the disk and occasionally in emission (\citeads{1966soat.book.....Z}; \citeads{1973SoPh...28...95R}; \citeads{1988assu.book.....Z}). Contrary to sprays, the motion of the material is ordered and appears to follow the magnetic field lines of force, while often the material comes back along the same path.

\begin{figure}[h]
\centering
\includegraphics[width=\hsize]{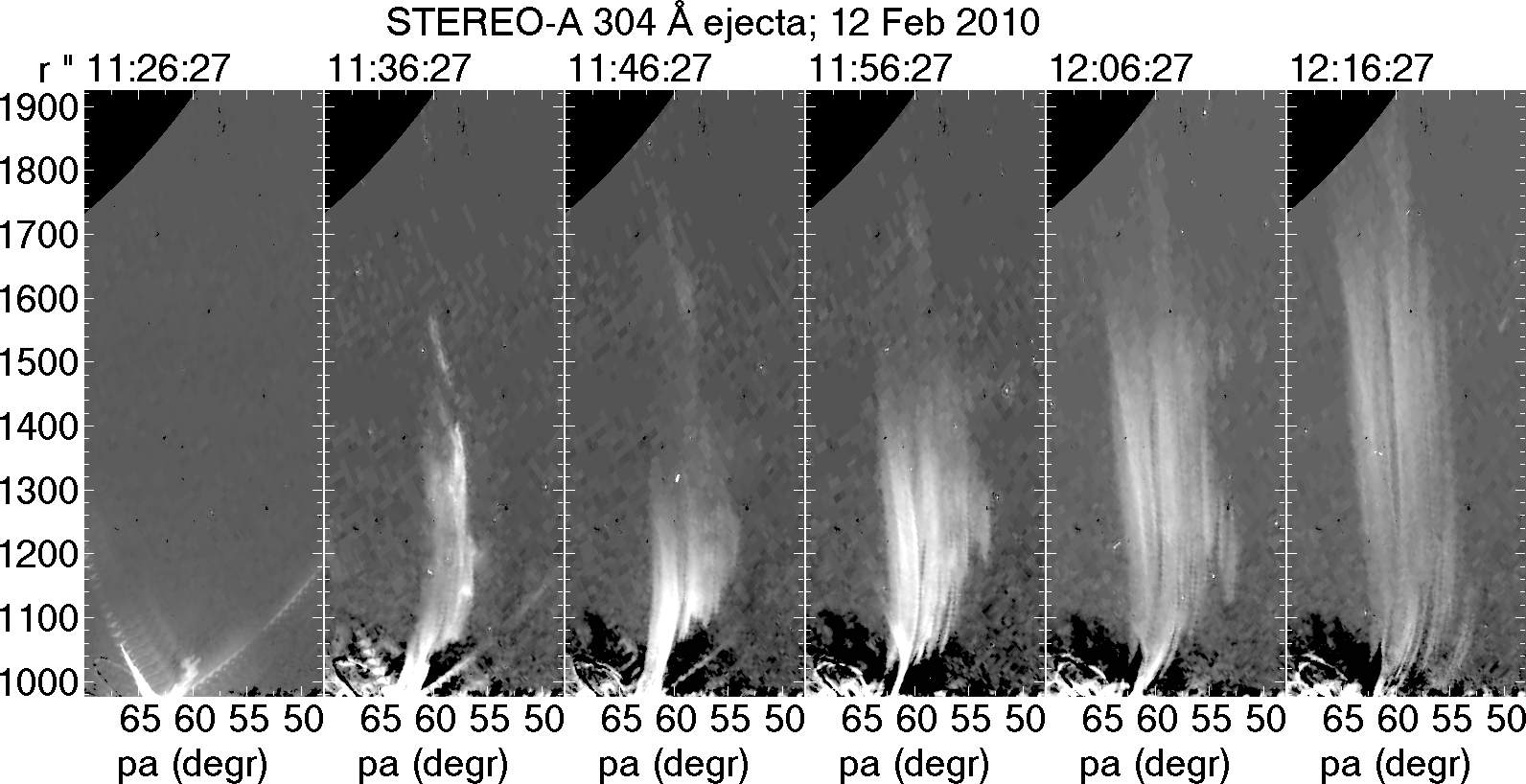}
\caption{Base difference images of the surge in the 304\,\AA\ band, as seen from STEREO-A. The intensity is given as a function of position angle and distance from the center of the disk. The solar radius is 972\arcsec. }
\label{304surge}
\end{figure}

The development of soft X-ray, ultraviolet (UV), and Extreme Ultraviolet (EUV) instrumentation has helped us identify ejecta in these wavelength ranges as well. They are usually called jets and their properties have been reviewed by \citetads{2016SSRv..201....1R}. It appears that in several cases the ejected material is multi-thermal which allows the detection of its cool, dense component in \ha\ as a surge and the detection of its hotter, more tenuous component in (E)UV or soft X-rays as a jet. Using Interface Region Imaging Spectrograph (IRIS) data and numerical simulations, \citetads{2021A&A...655A..28N} have found that surges are characterized by temperatures around 6000\,K at $-6.0 \leq \log_{10} (\tau) \leq -3.2$ and electron density ranging from $1.6 \times 10^{11}$ up to about 10$^{12}$\,cm$^{-3}$ at $-6.0 \leq \log_{10} (\tau) \leq -4.8$. On the other hand, in soft X-rays and the EUV the temperature of jets may exceed 2\,MK while their density can reach values as high as 10$^9$-10$^{10}$\,cm$^{-3}$ (\citeads[e.g.~]{2016A&A...589A..79M}). Atmospheric Imaging Assembly (AIA) observations of a surge at 304 \AA\ yielded an average density of 4.1$\times$10$^{9}$\,cm$^{-3}$ \citepads{2013ApJ...770L...3K}.

There is a long tradition of interpreting jets and surges as a result of magnetic reconnection between newly emerged magnetic flux and the preexisting coronal magnetic field (\citeads[e.g.~]{1992PASJ...44L.173S}; Yokoyama \& Shibata \citeyearads{1995Natur.375...42Y},\citeyearads{1996PASJ...48..353Y}; Archontis \& Hood \citeyearads{2008ApJ...674L.113A}, \citeyearads{2012A&A...537A..62A}, \citeyearads{2013ApJ...769L..21A}; \citeads{2013ApJ...771...20M}). Observations of jets associated with the eruption of small-scale filaments (e.g. Sterling et al. \citeyearads{2015Natur.523..437S}, \citeyearads{2016ApJ...821..100S}) has led \citetads[][\citeyearads{2018ApJ...852...98W}]{2017Natur.544..452W} to propose that the ejected material is generated by a break-out mechanism similar to the one that might be at work in the initiation of CMEs (\citeads{1999ApJ...510..485A}).

\begin{figure}[h]
\centering
\includegraphics[width=\hsize]{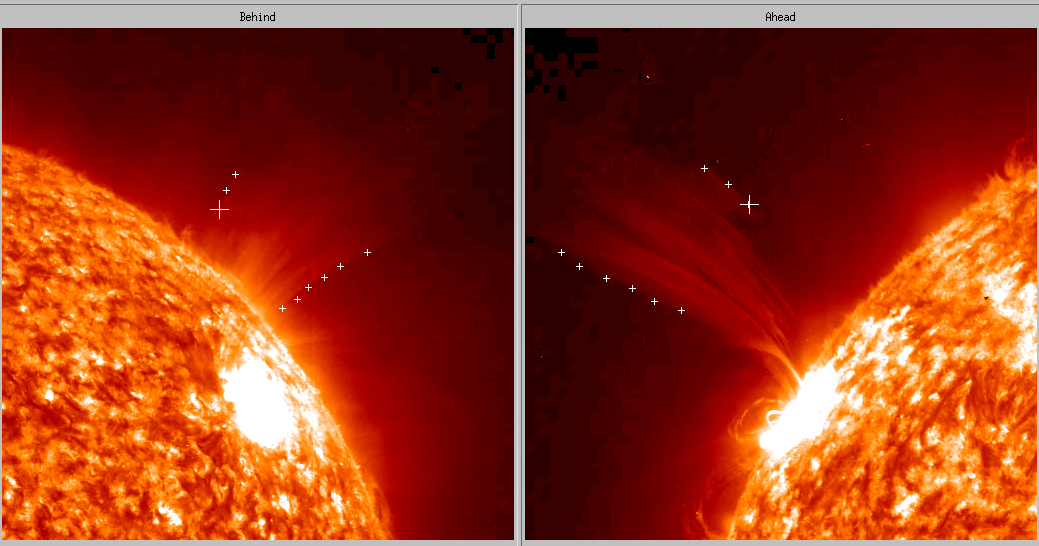}
\caption{
STEREO-B (left) and A (right) images of the surge in the 304\,\AA\ band at 11:56 UT. The crosses mark the triangulation points}
\label{304triang}
\end{figure}

In a previous work (\citeads{2021A&A...654A.112A}, hereafter Paper I), we presented a detailed analysis of the complex M9 event SOL2010-02-012T11:25:00, whose dynamic spectrum contained at least five lanes of type-II emission and included a surge, a CME and an extreme ultraviolet (EUV) wave. Our analysis showed that the type IIs were not associated with the surge, but rather with the EUV wave probably initiated by the surge. During the main phase of the event there was no detectable metric emission from the surge. 

In this article we report the detection of surge-associated emission, which occurred after the end of the bulk of the metric radio burst. In the next section we present our observations and in Sect.~\ref{results} our results. In Sect.~\ref{emission} we discuss possible emission mechanisms, whereas in Sect.~\ref{concl} we present our conclusions. Finally in Appendix \ref{Appendix} we give the full dynamic spectrum of the event and a movie of images used in this study.

\begin{figure*}[h]
\centering
\includegraphics[width=.96\hsize]{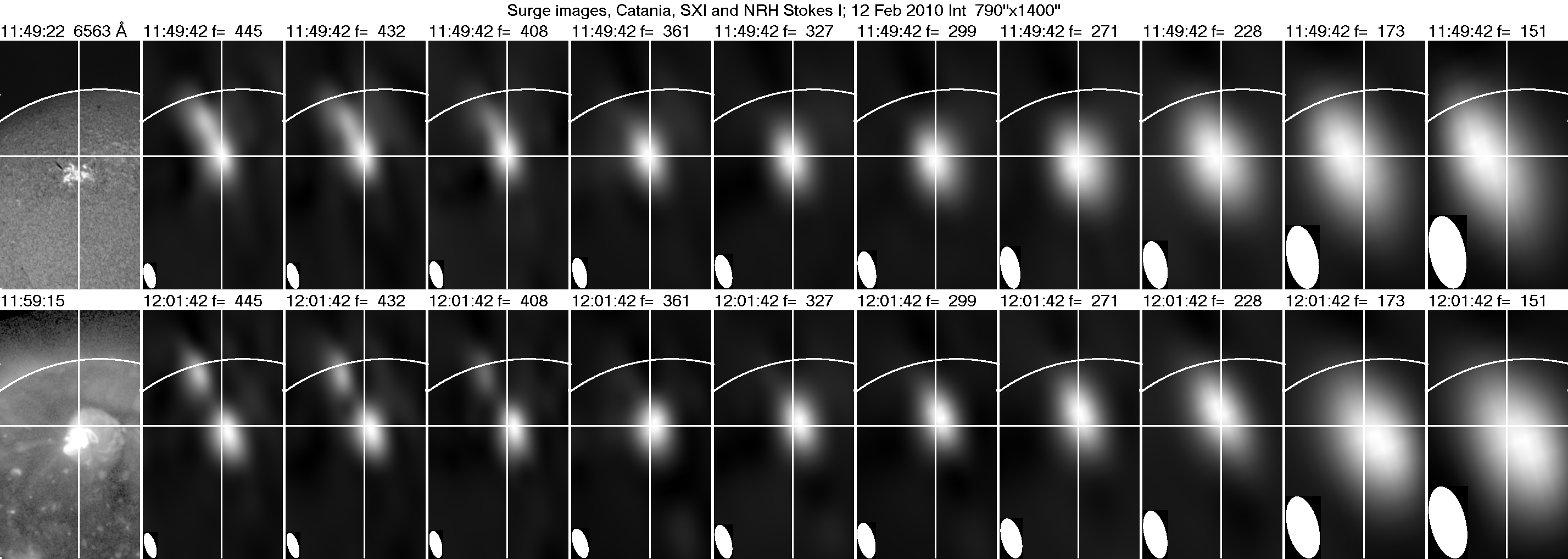}
\caption{NRH images in total intensity (Stokes parameter I) of the surge at all 10 frequencies during the first (top) and the second (bottom) phase of the event. The white arc marks the solar limb, the insert shows the NRH beam (resolution), while the cross-hair is placed at the maximum of the lower source at 445\,MHz during the first phase.  { The images are normalized so that the maximum intensity at each frequency is white and the minimum is black, with a linear scale ($\gamma=1$.).} An \ha\ image from Catania and a long exposure SXR image from SXI are given for reference in the left column.}
\label{NRH_Cat_SXIw}
\end{figure*}

\begin{figure*}[h]
\centering
\includegraphics[width=.96\hsize]{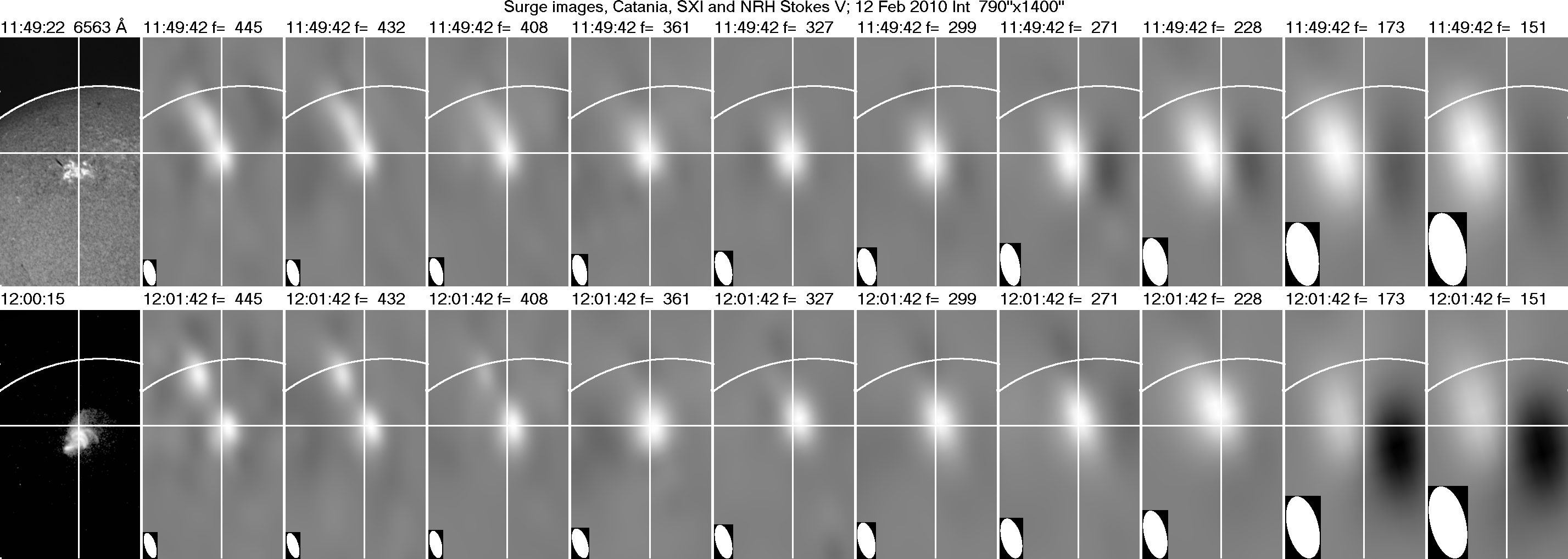}
\caption{Same as Fig.~\ref{NRH_Cat_SXIw} for circular polarization (Stokes V). Here a short exposure SXI image is given, to show better the structure of the active region { and zero V is gray}.}
\label{NRH_Cat_SXIw_V}
\end{figure*}

\section{Observations and data analysis}
The data set used in this work is similar to the one used for Paper I, for the time interval 11:45 to 13:00 UT of February 12, 2010. In short, we used images from the Nan\c cay Radioheliograph ({\citeads[NRH, ]{1997LNP...483..192K}})\footnote{downloaded from http://secchirh.obspm.fr/}, dynamic spectra from the ARTEMIS-JLS radiospectrograph ({\citeads{2001ExA....11...23C}, \citeads{2006ExA....21...41K}}) and CALLISTO\footnote{downloaded from http://soleil.i4ds.ch/solarradio/callistoQuicklooks/}, light curves from the Geostationary Operational Environmental Satellite (GOES), \ha\ images from Catania Observatory, soft X-ray images from the GOES {\it Soft X-ray Imager} (SXI), EUV images at 304 \AA\ with a cadence of 10 min from STEREO {\it Sun-Earth Connection Coronal and Heliospheric Investigation} (SECCHI) Ahead and Behind and coronograph images from STEREO Coronagraph1 (COR1) A and B (\citeads{2008SSRv..136...67H}) with a cadence of 5 min. No Solar and Heliospheric Observatory (SOHO) {\it Extreme ultraviolet Imaging Telescope} (EIT) data were available during the event, while the {\it Transition Region and Coronal Explorer} (TRACE) was pointed elsewhere.

The NRH provided data at ten frequencies (150.9, 173.2, 228.0, 270.6, 298.7, 327.0, 360.8, 408.0, 432.0 and 445.5\,MHz) with a cadence of 250 ms and a spatial resolution of 1.2\arcmin\ by 1.8\arcmin\ at 432 MHz. For a global view of the event we used 10\,s average images, whereas for a detailed study we used images with the full temporal resolution. To facilitate displays and comparisons, we also computed cuts of intensity as a function of time and position, along the axis of the surge. Three-dimensional positions were computed from triangulation of STEREO A and B 304\,\AA\ images, as well as from COR1 images.

\section{Results}\label{results}
\subsection{Overview}
The GOES class M9 event, which occurred around 11:25 UT on February 12, 2010 in active region 11046, was described in detail in Paper I. The bulk of the radio emission ended around 11:32 UT (see Fig. 1 of Paper I), but metric continuum emission appeared again in the dynamic spectrum from { about} 12:00 to { 13:00} UT (see Fig. 45 of \citeads{2015SoPh..290..219B}, { reproduced in Fig.~\ref{FullDS}}){; this continuum is associated with the second phase of the surge described in this work}. 

In the STEREO images the surge started around 11:23 UT and was visible in the 304\,\AA\ band well after the main metric emission, till about 13:26 UT. In the 195\,\AA\ band it was visible in emission up to about 11:38 UT, whereas the top of the surge went above the COR1 A and B occulting disks at around 11:40 UT 

Seen from STEREO-A, the event was very close to the E limb, whereas it was near the W limb for STEREO-B { (Fig.~\ref{STEREOpos})}. Fig.~\ref{304surge} shows a sequence of STEREO-A images, in which the surge is very prominent as it is seen mostly beyond the limb. We note that between 11:36 and 11:46 UT a second ejection of material occurred, as noted in Paper I (Sect. 3.2 and Fig. 5) where it was characterized as a ``slow component''. For comparison, we also show a pair of STEREO A and B 304\,\AA\ images in Fig.~\ref{304triang}, where the points used for triangulation are marked.

Radio emission from the surge was detected by the NRH during two intervals: from 11:49 to 11:51 UT (phase A) and from 12:00 to $\sim$13:00 UT (phase B). Fig.~\ref{NRH_Cat_SXIw} shows NRH total intensity images during both phases, together with \ha\ and SXR images. The corresponding circular polarization images are shown in Fig.~\ref{NRH_Cat_SXIw_V}.

We first note that there is no trace of surge emission either in \ha\ or in SXR. We further note that in both phases two sources appear in the high frequency images where the resolution is higher; moreover, during phase B the upper source is clearly displaced  upwards, compared to the phase A, apparently a result of the expansion of the surge. A final remark at this stage is that the size and position of the source are markedly different in the two lowest frequencies, indicating different structures and/or different emission mechanisms. This is further corroborated by inspection of the V images, which show a bipolar structure at low frequencies, more prominent during phase B, whereas in high frequencies both the lower and the upper source were uniformly polarized in the right-hand sense. 

\subsection{Low time resolution analysis}\label{LowResAn}
At the full cadence (250\,ms) of the NRH, our 65\,min of data at 10 frequencies and two Stokes parameters make a total of 312\,000 images. This is a huge number of images, which cannot be easily treated. For this reason we will start our analysis with 10\,s average images which, while preserving the basic characteristics of the event, reduced the number of images by a factor of 40. 

\subsubsection{Evolution in time}\label{Evol}
Even at the reduced time resolution, the number of images is still large. A convenient way of presentation is through cuts of the intensity as a function of position and time, which are displayed in Fig.~\ref{CutsI_LowResW}. The full set of averaged images is presented in Movie 1 of the Appendix (see Fig.~\ref{frames}). 

\begin{figure}[t]
\centering
\includegraphics[width=.98\hsize]{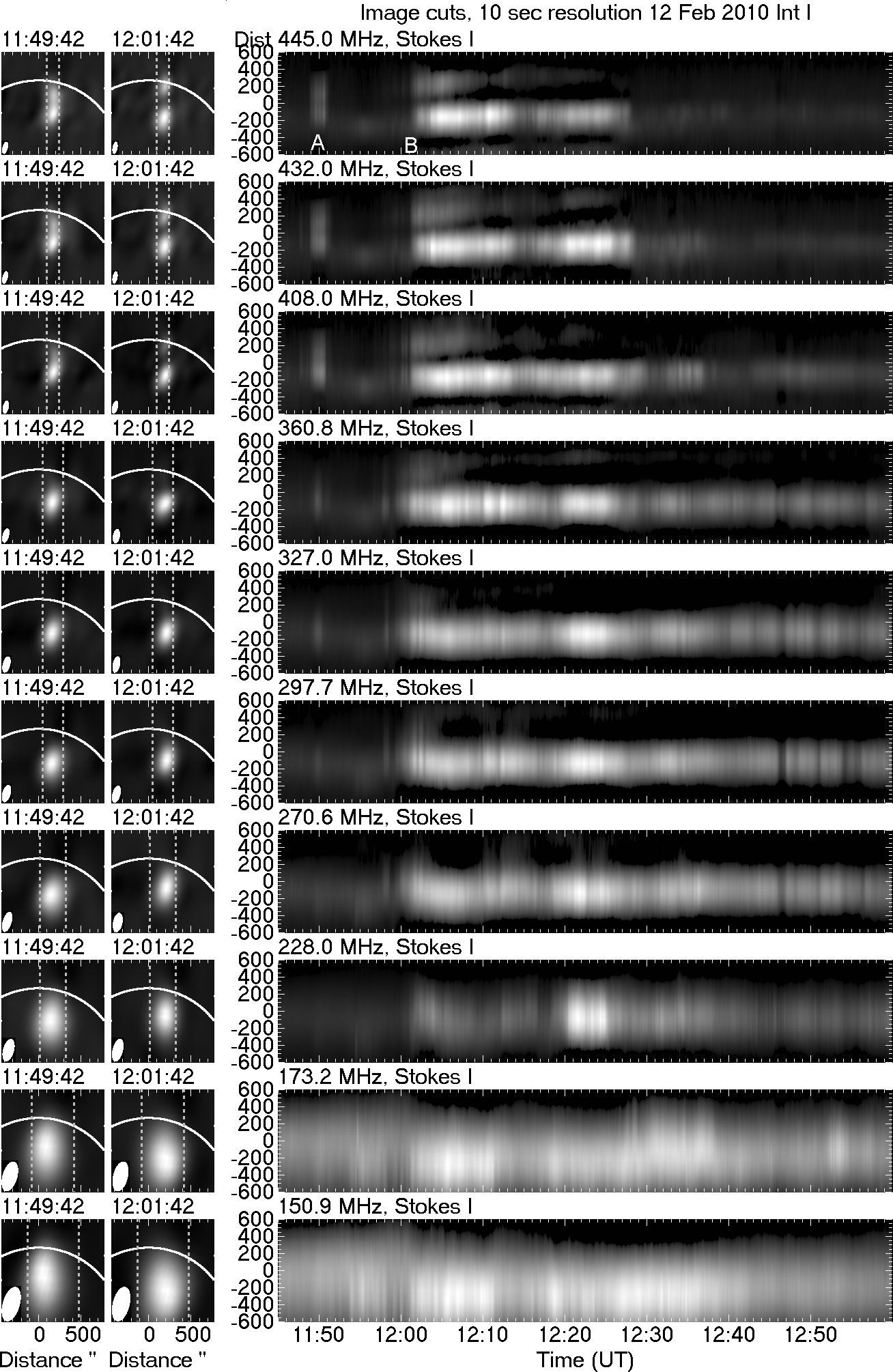}
\caption{Cuts of intensity as a function of time and position along the surge axis from 10\,s resolution NRH total intensity images. Zero position is at the middle of the cut. { Corresponding} images during the first phase and the beginning of the second phase of the surge are given on the left of the cuts; dashed vertical lines mark the limits of the cut region. The width of the cut changes with frequency in order to cover the entire emitting region. { A and B in the 445.0\,MHz cut mark the two phases of the event.} We note that { phase A was} much shorter and weaker than { phase B}.}
\label{CutsI_LowResW}
\end{figure}

The cuts confirm our previous conclusion that the nature of the source is different at the two low frequencies; indeed, the cuts at 173.2 and 150.9 MHz show fluctuating emission but no concrete sign of the two phases. Moreover, the cuts provide some additional interesting information:
 
The emission during phase B was much brighter and longer than the emission during phase A. In addition, the duration of this phase increased as the frequency decreased, extending beyond 13:00 UT. The emission from the lower source during the same phase can roughly be split in three parts, best visible at high frequencies: from 12:00 to 12:15 UT, from 12:15 UT to 12:28 UT and beyond 12:28 UT.

The visibility of the upper source decreased with decreasing frequency. In order to check if this is due to the NRH resolution, we convolved  the 445\,MHz image with the beam of all other frequencies, after deconvolving with the 445\,MHz beam. We found that the upper source should be detectable even at 150.9\,MHz, during phase B in particular where the two sources are further apart. This means that the intensity of the upper source, relative to the intensity of the lower source, increases with frequency. 

\begin{figure}[]
\centering
\includegraphics[width=.9\hsize]{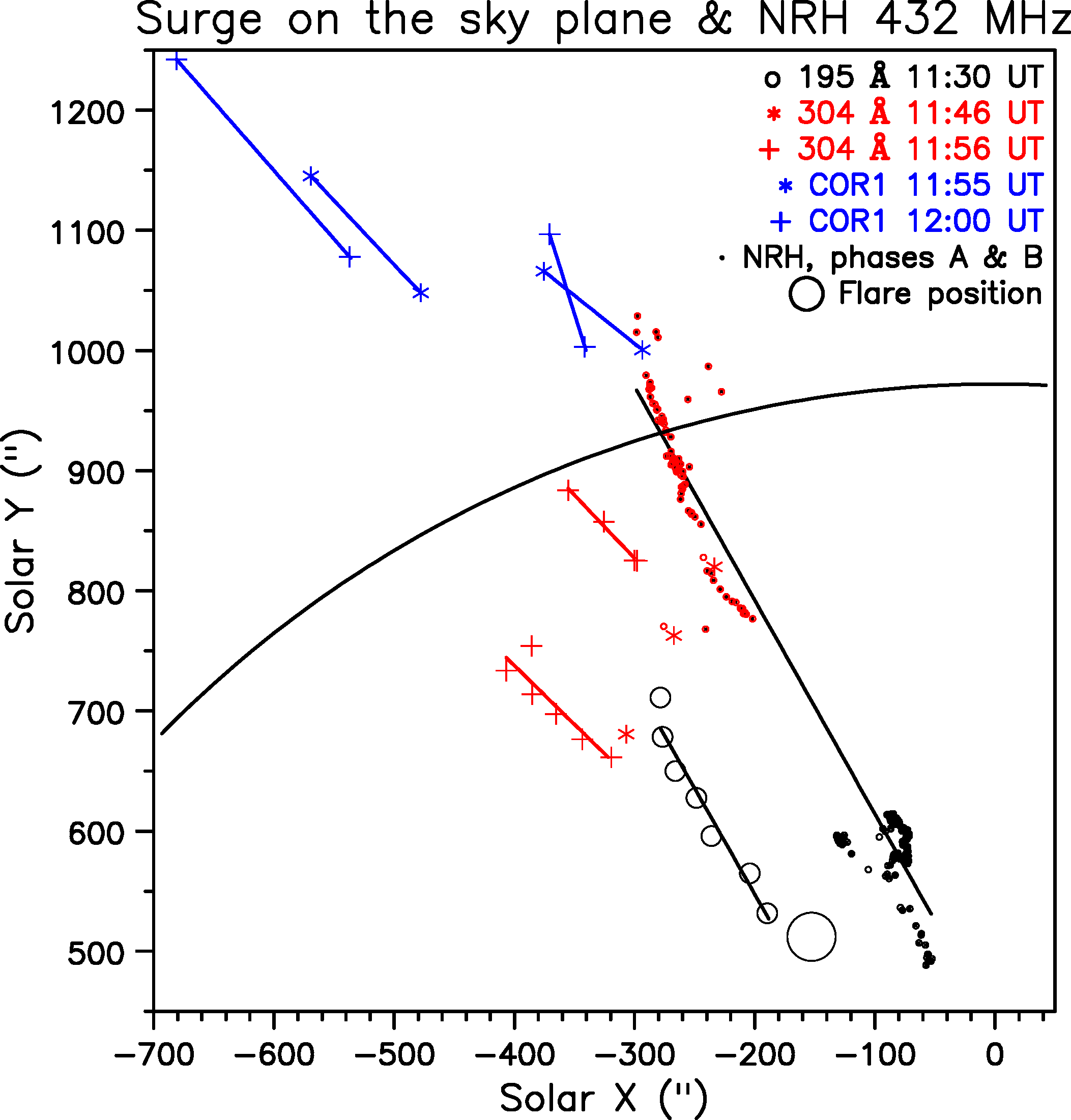}
\caption{Positions of various components of the surge derived from triangulation of STEREO images on the plane of the sky as seen from the Earth. The position of the radio sources at 432.0\,MHz are shown as black and red dots, for the lower and upper components of the emission respectively. The open circle shows the location of the flare. Straight lines are from linear regression. The black arc marks the solar limb.}
\label{Orient}
\end{figure}

\subsubsection{Positions and motions}
As mentioned above, the three-dimensional position of surge features were computed from triangulation of STEREO A and B images. From these, the projection of the surge on the plane of the sky, as seen from the Earth, was computed and is displayed in  Fig.~\ref{Orient}. The position (centroid), as well as the brightness temperature and size of the radio sources was computed by Gaussian fit of their images. The positions derived from 10\,s average images at 432.0\,MHz during the entire event are plotted as dots in Fig.~\ref{Orient}, black for the lower source and red for the upper. The flare position, from XRT and Catania images, is shown as an open circle in the same figure (see also Fig.~6 in Paper I).

A first remark is that the surge was huge, its projected length being about a solar radius around 12:00\,UT. On the average, its position angle was about 34\degr\ east of north. The radio source positions were spread along a shorter, but still quite long structure, going as far as the low part of COR1A emission. The extent of radio emission was about 0.5\,R$_\sun$, and was displaced by about 150\,\arcsec\ to the west of the surge; this displacement is comparable to the observed width of the sources and might be due to refraction/scattering effects or triangulation errors. We add that the orientation of the radio source motion was close to that of the surge, about 29\degr.

From the position of the top of the surge in STEREO-A 304\,\AA\ images, we deduced an accelerating motion and we obtained a velocity, projected on the plane of the sky, of about 126\,km\,s$^{-1}$ around 12:00\,UT; this is consistent with the value of 120\,km\,s$^{-1}$, obtained from COR1A measurements in Paper I (Sect. 3.2) for this component of the surge. As, according to our triangulation the surge was inclined by about 15\degr\ with respect to the plane of the sky as seen from STEREO-A, this velocity translates to 130\,km\,s$^{-1}$ along the surge direction. On the other hand, a linear regression of the positions of the upper NRH source at 445.0\,MHz as a function of time gave an apparent velocity of 123\,km\,s$^{-1}$; seen from the Earth, the inclination of  the surge with respect to the plane of the sky was 45\degr (see Fig.~\ref{Inclin} below), thus this value corresponds to 175\,km\,s$^{-1}$ along the surge. Although this value is $\sim$35\% above the one deduced from the limb measurements, it is probably more reliable since the source centroids can be more accurately identified and measured than the defuse top of the 304\,\AA\ images. { We note that the upward displacement of the lower source (black dots in Fig.~\ref{Orient}) is considerably smaller ($\simeq40$\,km\,s$^{-1}$) than that of the upper source (red dots); this is expected, since the material in the lower source is continuously replenished from the eruption that produced the surge.}  
 
\begin{figure}[h]
\centering
\includegraphics[width=.9\hsize]{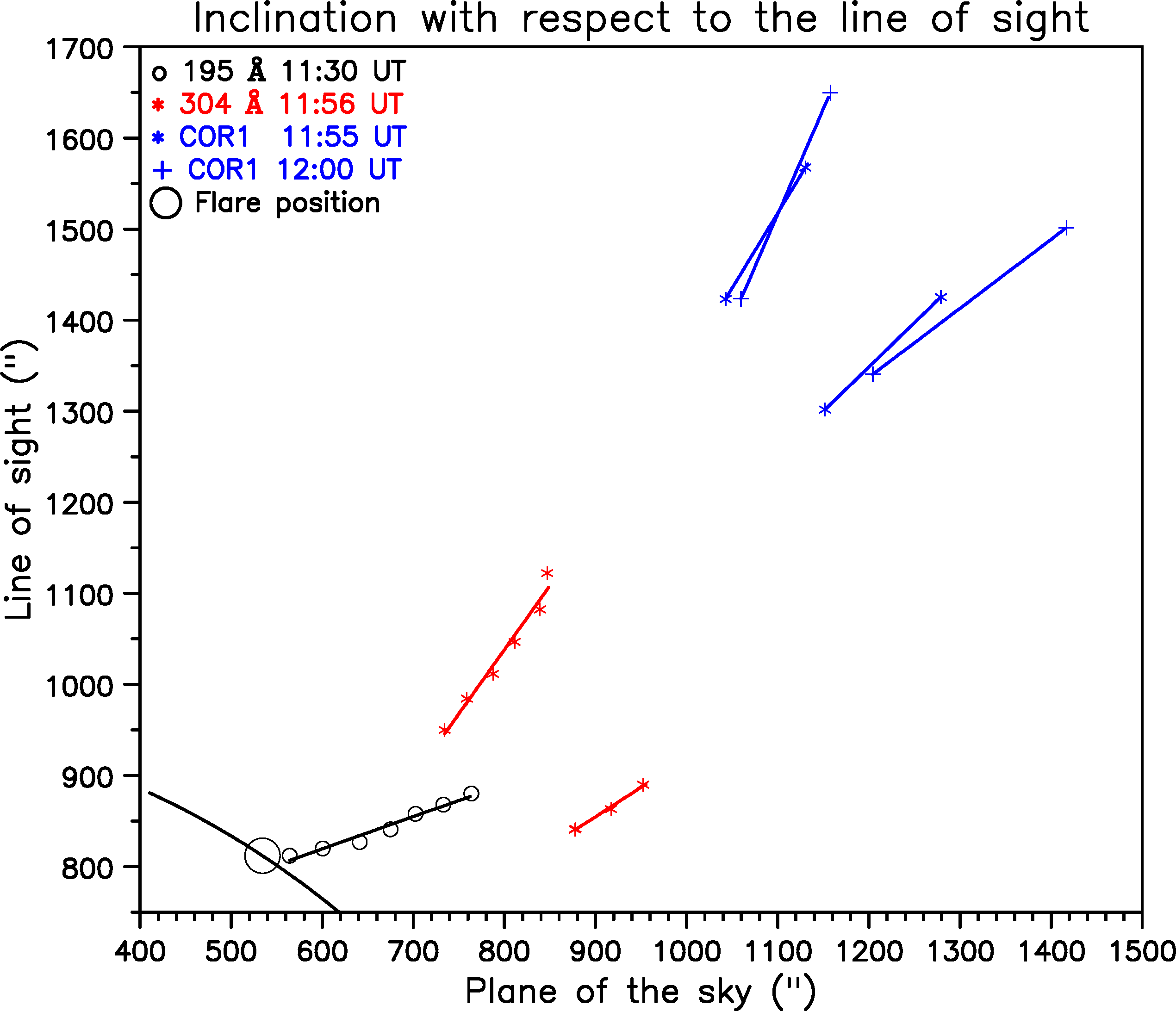}
\caption{Positions of the surge components derived from triangulation of STEREO images on a plane perpendicular to the plane of the sky. The open circle shows the location of the flare. Straight lines are from linear regression. The black arc marks the solar limb.}
\label{Inclin}
\end{figure}

Putting together all information presented in this section, there can be no doubt that what we see from the Earth at metric wavelengths is indeed emission associated with the 304\,\AA\ surge seen near the limb from STEREO A and B.

In order to complete the picture, we provide in Fig.~\ref{Inclin} the surge positions, derived from triangulation, projected on a plane that includes the line of sight. Apart from providing the inclination of the surge with respect to the plane of the sky as seen from the Earth, the plot shows a change of the surge inclination with height, apparently reflecting the curved shape of this structure seen in Fig.~\ref{304surge}.

\subsection{Intensity fluctuations from full time resolution data}\label{fluct}
Although our 10\,s average images give an accurate view of the structure and the evolution of the event, there is a lot more of information in the full time sequence if NRH images, which will be exploited in this section, separately for phases A and B.

\subsubsection{Phase A}\label{PhaseA}
\begin{figure}[h]
\centering
\includegraphics[width=\hsize]{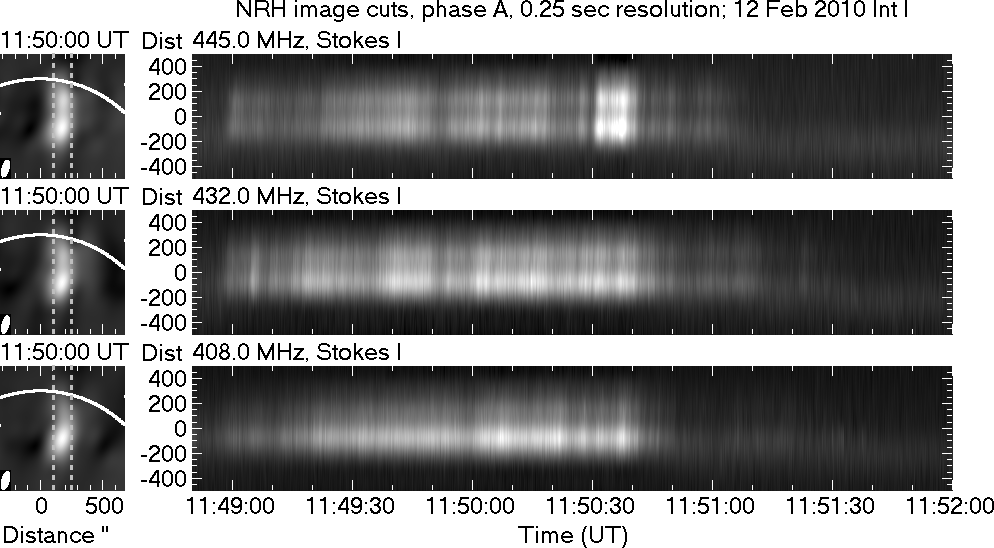}
\caption{Stokes $I$ cuts (intensity as a function of time and position along the surge), at full time resolution, for 445.0, 432.0 and 408.0\,MHz, during phase A of the surge. Zero position is at the middle of the cut. The width of the cuts is marked on the images at left.}
\label{cutsA}
\end{figure}

The first phase of the metric surge emission was rather short, lasting for about 2\,min. A concise picture of its structure and evolution is given in the total intensity cuts (Fig.~\ref{cutsA}) at the three highest NRH frequencies, where the two sources were clearly resolved; the circular polarization cuts show a similar picture and are not shown here.

The time variation of the cuts of Fig.~\ref{cutsA} reveals two temporal components of the emission: one with a time scale of the order of 1\,min and a second, fluctuating component, with sub-second scale. A power spectrum analysis did not reveal any concrete periodicities in the fluctuations, which occurred at an average rate of about one per second.

\begin{figure}[]
\centering
\includegraphics[width=\hsize]{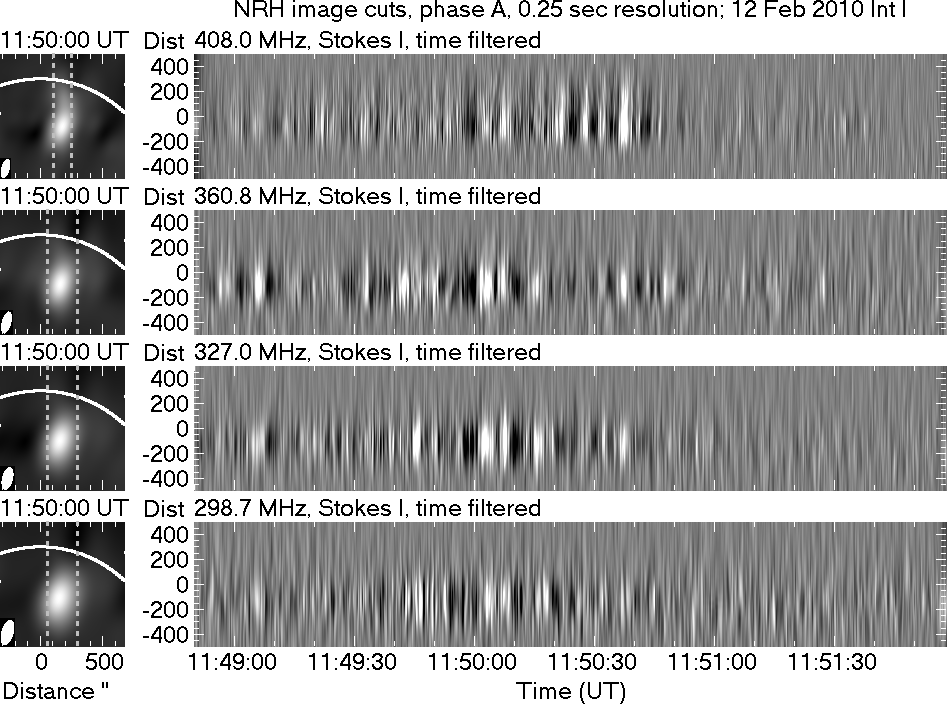}
\caption{High-pass time filtered Stokes $I$ cuts, for 408.0, 360.8, 327.0 and 298,7\,MHz. The width of the cuts is marked on the images at left.}
\label{cutsFilt}
\end{figure}

The long time-scale component is broad-band, detectable from 445.0\,MHz down to about 298.7\,MHz, whereas most fluctuations are very different in the consecutive NRH channels displayed in the figure, which puts an upper limit of a few tens of MHz in their bandwidth. Still, there are some short time scale structures in the 408.0-298.7\,MHz frequency range, which are visible in more than one NRH channel, suggesting a bandwidth of the order of 100\,MHz; this is shown in Fig.~\ref{cutsFilt}, where a high-pass Gaussian filter of 5\,s width has been applied to the cuts in order to enhance fine structures.

Most interesting, the pattern of fluctuations is very similar in the two sources, with a slight time delay of 0.2\,s of the upper source with respect to the lower, as measured through cross-correlation of their light curves; this appears as a small inclination in the cuts of Fig.~\ref{cutsA}. Taken literally, this translates to an apparent superluminal propagation speed of 2.6c on the plane of the sky, the projected distance of the two sources being about 230\arcsec. As the geometry has to be taken properly into consideration, we will come back to this issue in Sect.~\ref{emission}.
  
\begin{figure}[h]
\centering
\includegraphics[width=\hsize]{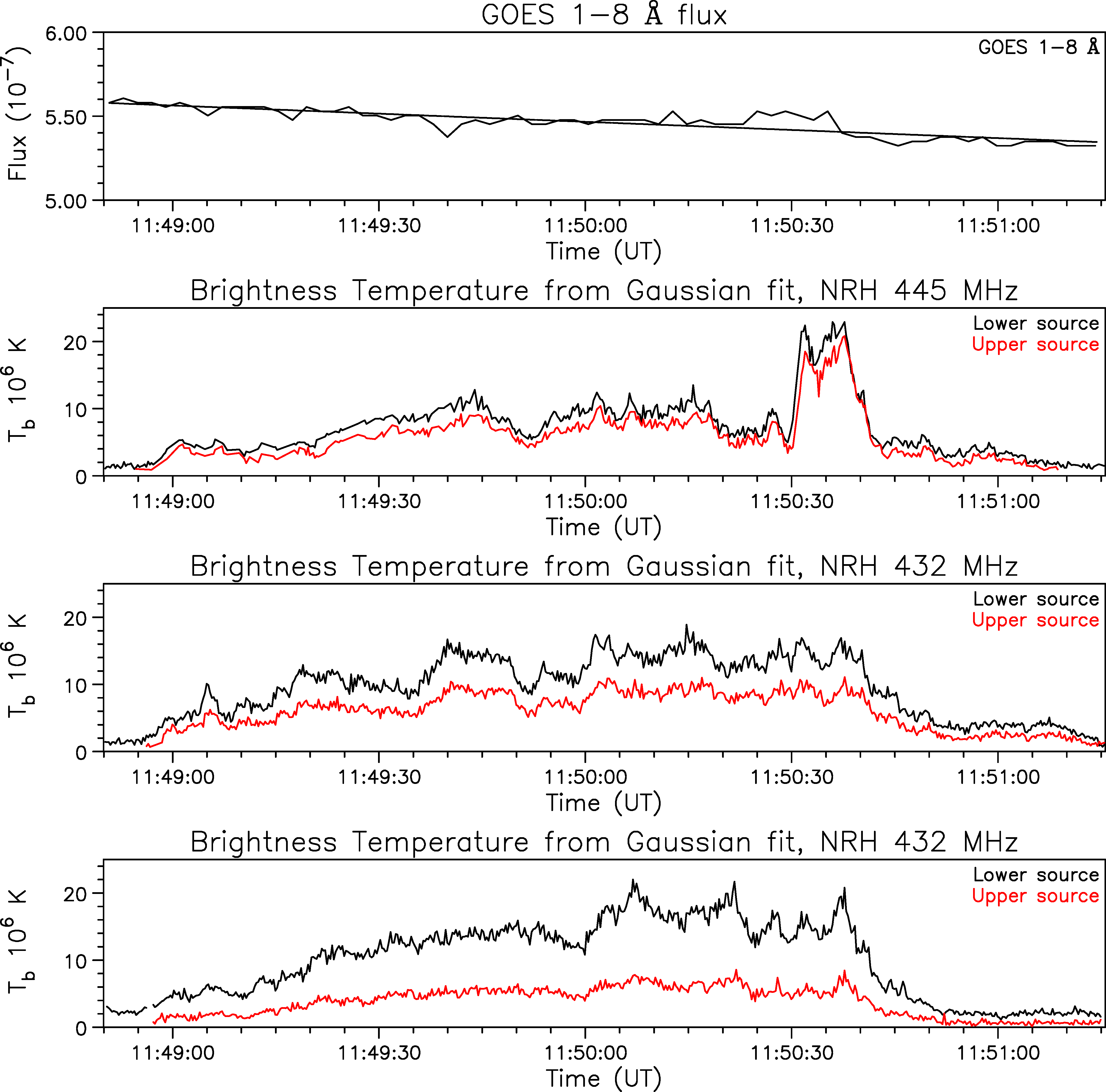}
\caption{Light curves of brightness temperature for the three frequencies shown in Fig,~\ref{cutsA}. The GOES light curve is also shown for reference, where the straight line marks the linear fit of the background.}
\label{lightA}
\end{figure}

As expected, the light curves of the two sources follow closely each other for all three frequencies considered here (Fig.~\ref{lightA}). The time shift between the two sources is also visible in this figure and so is the narrow bandwidth of the fluctuations. The brightness temperature, $T_b$, of the lower source is of the order of 10$^7$\,K. At 445.0\,MHz the brightness of the upper source is slightly lower than that of the lower source, their ratio being 0.85; this value decreases to 0.60 at 432.0\,MHz and to 0.38 at 408.0\,MHz. This quantifies our previous remark (Sect.~\ref{Evol}) that the visibility of the upper source decreased with decreasing frequency. Finally, there appears to be no association of the radio emission to the GOES soft x-ray flux.

\subsubsection{Phase B}\label{PhaseB}
\begin{figure}[h]
\centering
\includegraphics[width=\hsize]{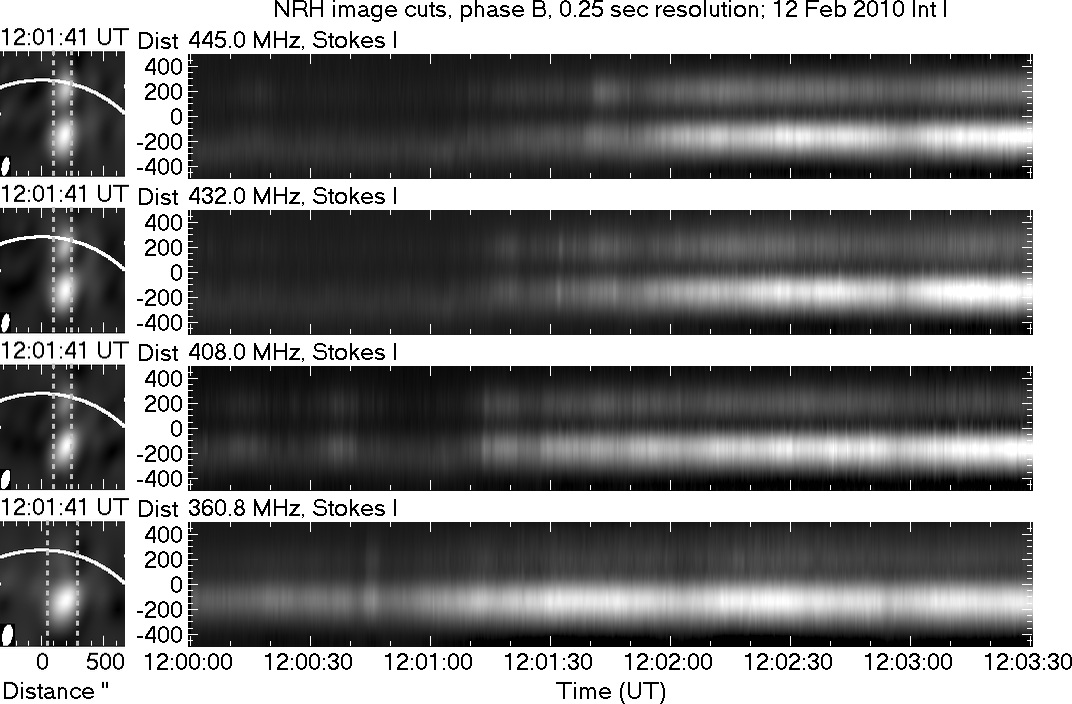}
\caption{Stokes $I$ cuts at full time resolution, for 445.0, 432.0, 408.0 and 360.8\,MHz, during phase B of the surge. Zero position is at the middle of the cut. The width of the cuts is marked on the images at left.}
\label{CutsB}
\end{figure}

As mentioned earlier, phase B was much longer and more intense than Phase A (c.f. Fig.~\ref{CutsI_LowResW}). The upper source is now detectable in the four higher NRH frequencies, fading as time progresses; it is further away from the lower source, a consequence of the expansion of the surge. As it is impossible to present its time evolution in full, we show in Fig.~\ref{CutsB} the first 3.5\,min.

As in Phase A, the intensity shows short time scale fluctuations; however, their coherence between the upper and lower sources is much reduced, particularly after 12:02 UT. Until then, the fluctuations in the upper source were delayed by 0.3\,s, 50\% more than during phase A; the increased delay is apparently related to the increased separation of the two sources.

\begin{figure}[h]
\centering
\includegraphics[width=\hsize]{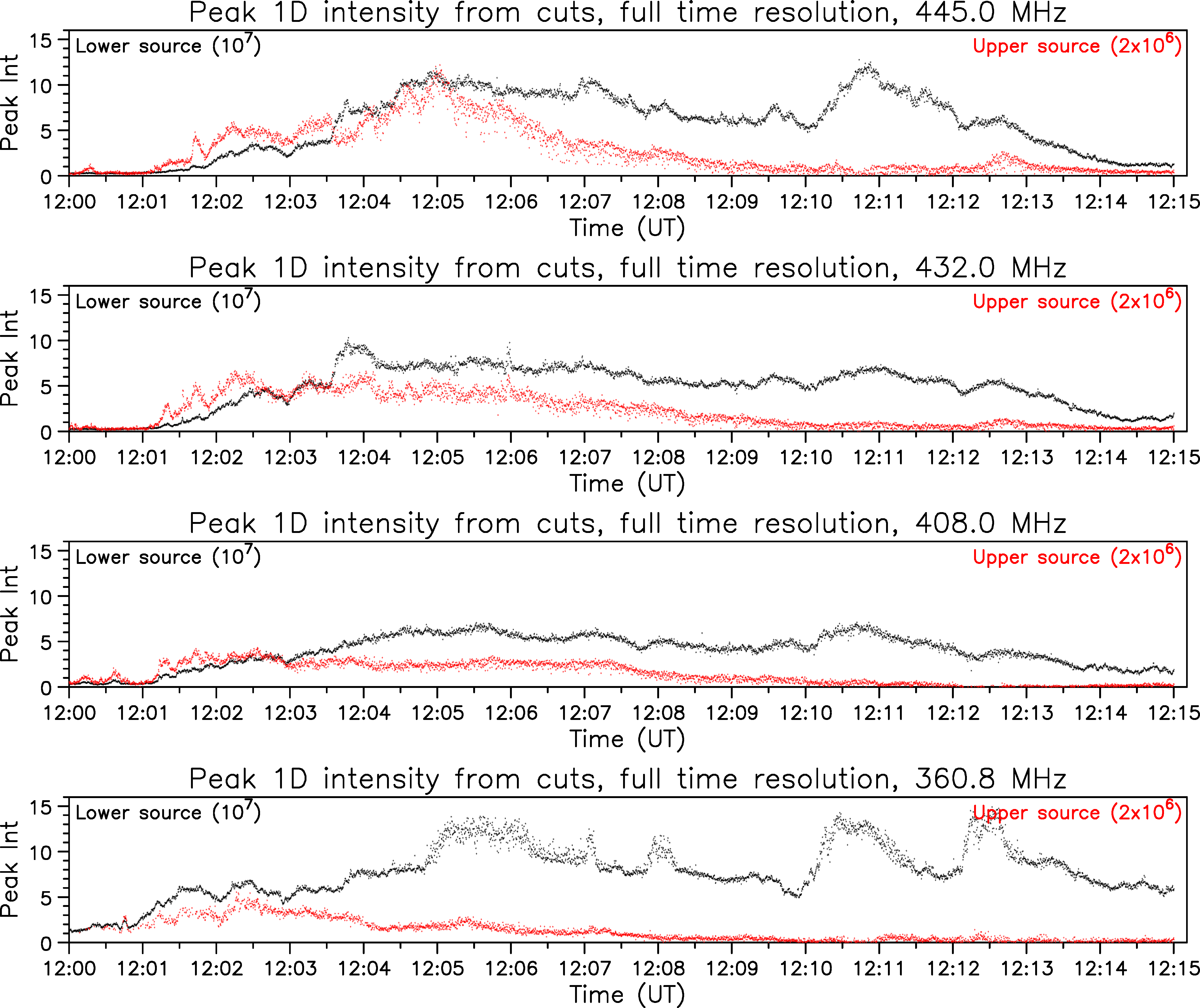}
\caption{Light curves at full time resolution, for 445.0, 432.0, 408.0 and 360.8\,MHz, during phase B of the surge. Intensity scales for the lower and upper sources are different by a factor of 5.}
\label{LightB}
\end{figure}

Fig.~\ref{LightB} compares the light curves of the upper and lower sources, computed from the 1D cuts. No GOES curve is given, as there was no SXR emission. Contrary to phase A, the intensity of the upper source not only is about a factor of 5 lower than that of the lower source, but it also decreases with time. This decrease could be associated with the increase of the distance between the two sources and/or the fading of the upper part of the surge.

\subsection{Spectrum and polarization}
\begin{figure}[h]
\centering
\includegraphics[width=\hsize]{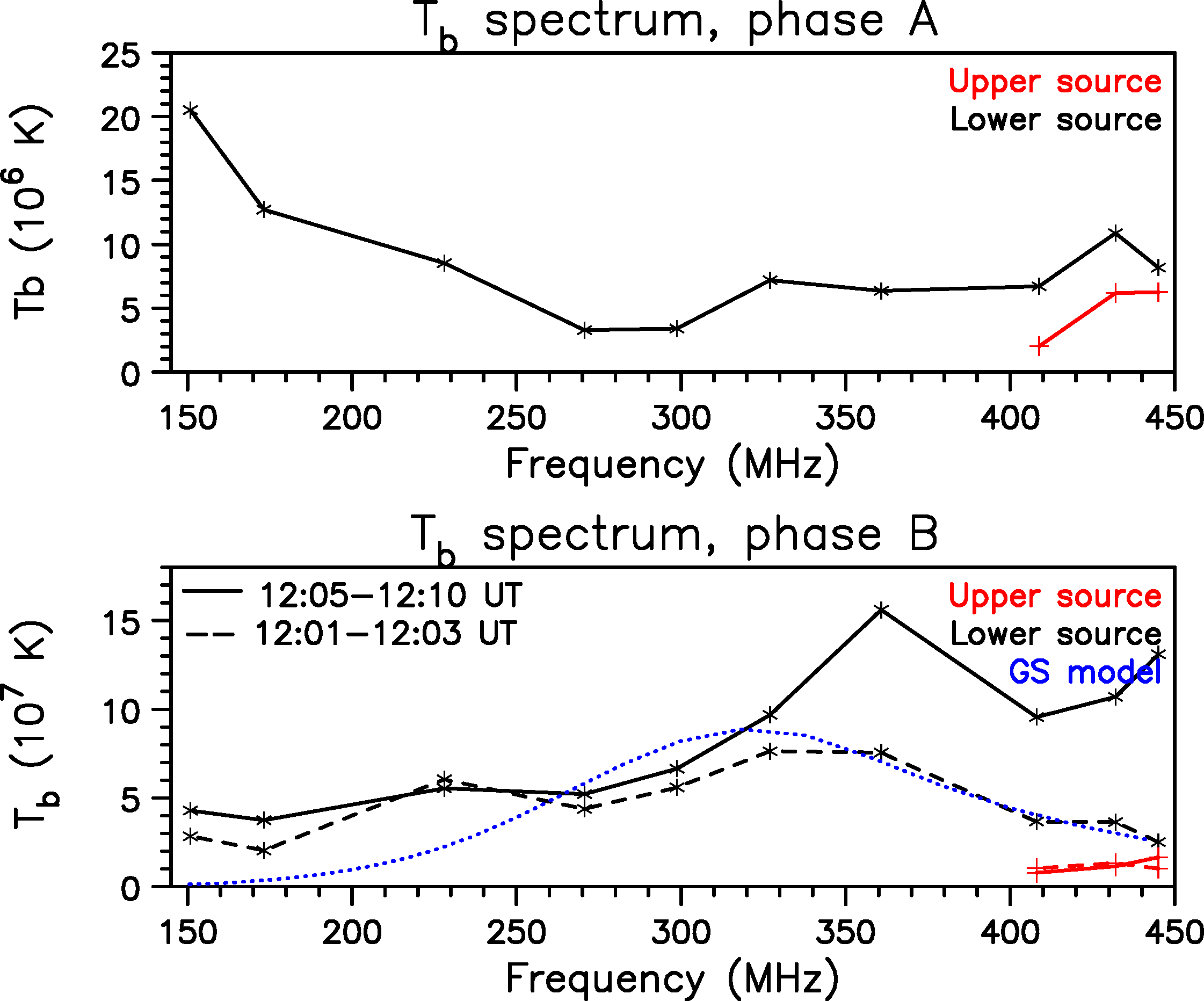}
\caption{Brightness temperature spectra during phase A (average, top panel) and during phase B (2 instances, bottom). The $T_b$ scales in the two panels differ by a factor of 10. The blue dotted curve in the lower panel shows a fit of the high frequency part of the spectrum with a gyrosynchrotron model discussed in Sect. \ref{mech}.}
\label{SpecAB}
\end{figure}

The average spectrum of the surge, computed from the peak NRH brightness temperature during phase A, is displayed in the upper panel of Fig.~\ref{SpecAB}. The low and high frequency spectral components discussed previously are clearly identifiable, with a cross-over frequency around 270\,MHz. We note that the high frequency component, which is the one directly associated with the burst, is rather flat with a peak around 432\,MHz ($T_b\simeq15$\,MK). 

During phase B both the low and the high frequency spectral components brightened, the latter more than the former; they are still distinguished in the spectrum (lower panel of Fig.~\ref{SpecAB}), better in the early stage of phase B (dashed line). The high frequency component now shows a prominent peak around 360\,MHz, whose brightness temperature exceeds 150\,MK. We note that in both phases the emission appears to extend above the highest NRH frequency of 445\,MHz.

\begin{figure}
\centering
\includegraphics[width=\hsize]{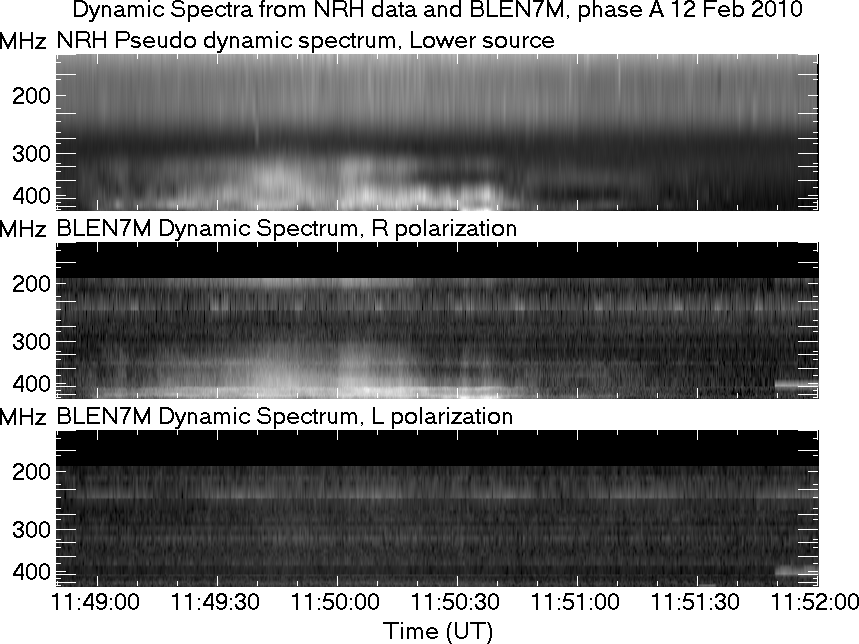}
\caption{Dynamic spectra during phase A of the surge from NRH imaging data (top) and from e-Callisto of Bleien Observatory in right and left circular polarization (middle and bottom) in the frequency range 150-525\,MHz. Horizontal line segments at left and at right mark the NRH frequencies. The time interval is the same as in Fig.~\ref{cutsA}.}
\label{DynSpecA}
\end{figure}

With 10 NRH channels between 150.9 and 445\,MHz, one could try to construct a dynamic spectrum that will give more complete spectro-temporal information. Such a ``pseudo dynamic spectrum" is presented in the top panel of Fig.~\ref{DynSpecA}. It compares well with the e-Callisto dynamic spectrum of Bleien Observatory (BLEN7M). The flux was below the detection threshold of other radiospectrographs, including ARTEMIS-JLS. Unfortunately the Bleien spectrum showed a sharp drop above 450\,MHz, apparently of instrumental origin as it appeared in all dynamic spectra from this instrument, and thus we can not estimate the high frequency extent of the emission.

\begin{figure}[h]
\centering
\includegraphics[width=\hsize]{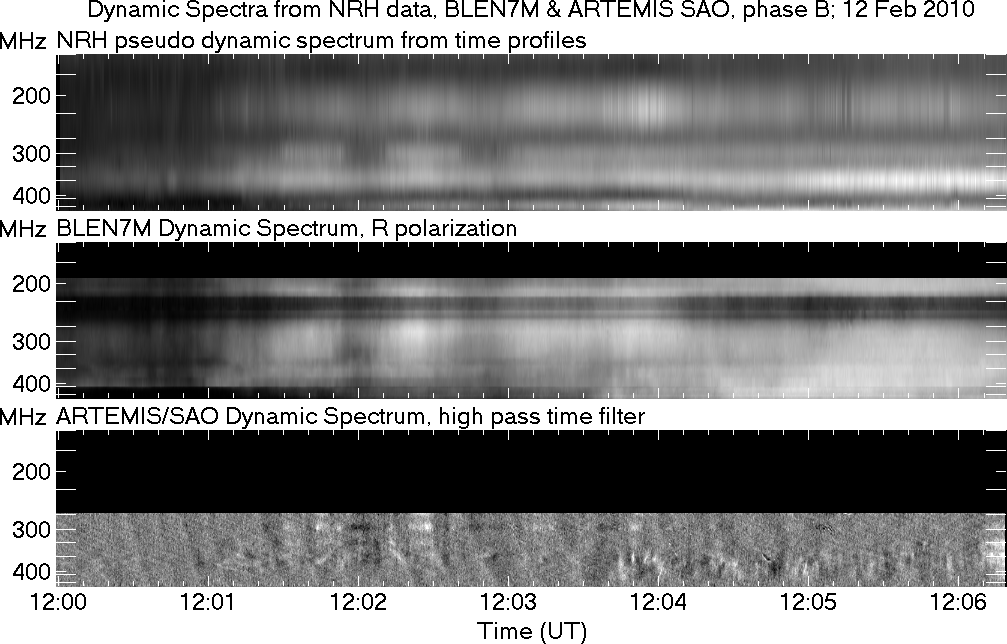}
\caption{Dynamic spectra from 12:00 to 12:06:20 UT, near the start of phase B of the surge from NRH imaging data (top), from e-Callisto of Bleien Observatory in right circular polarization (middle)  and from the acousto-optic analyzer of ARTEMIS-JLS (bottom, subjected to high pass time filtering). Fringe-like patterns in the bottom panel are of instrumental origin.}
\label{DynSpecB}
\end{figure}

Features of different temporal and frequency scales (c.f. Sect.~\ref{fluct}) are well visible in both spectra. In particular, repeated short duration spectral features visible down to $\sim300$\,MHz do not show any appreciable drift and they might be some sort of pulsations.

The BLEN7M spectra in Fig.~\ref{DynSpecA} reveal a very high degree of circular polarization in the right hand sense. This is confirmed by the NRH images, which gave a degree of polarization above 80\%.

Similarly, Fig.~\ref{DynSpecB} shows dynamic spectra during the second phase of the emission. In addition to the discrete low and high frequency components, these spectra clearly show pulsations, well visible in the BLEN7M spectrum of the middle panel. Moreover, after the application of a high pass time filter of 12\,s width, the spectrum obtained with the ARTEMIS-JLS acousto-optic analyzer (SAO) shows spike-like structures (bottom panel in the figure).

\section{Discussion}\label{emission}
\subsection{Apparent superluminal velocities}
\begin{figure}[h]
\centering
\includegraphics[width=\hsize]{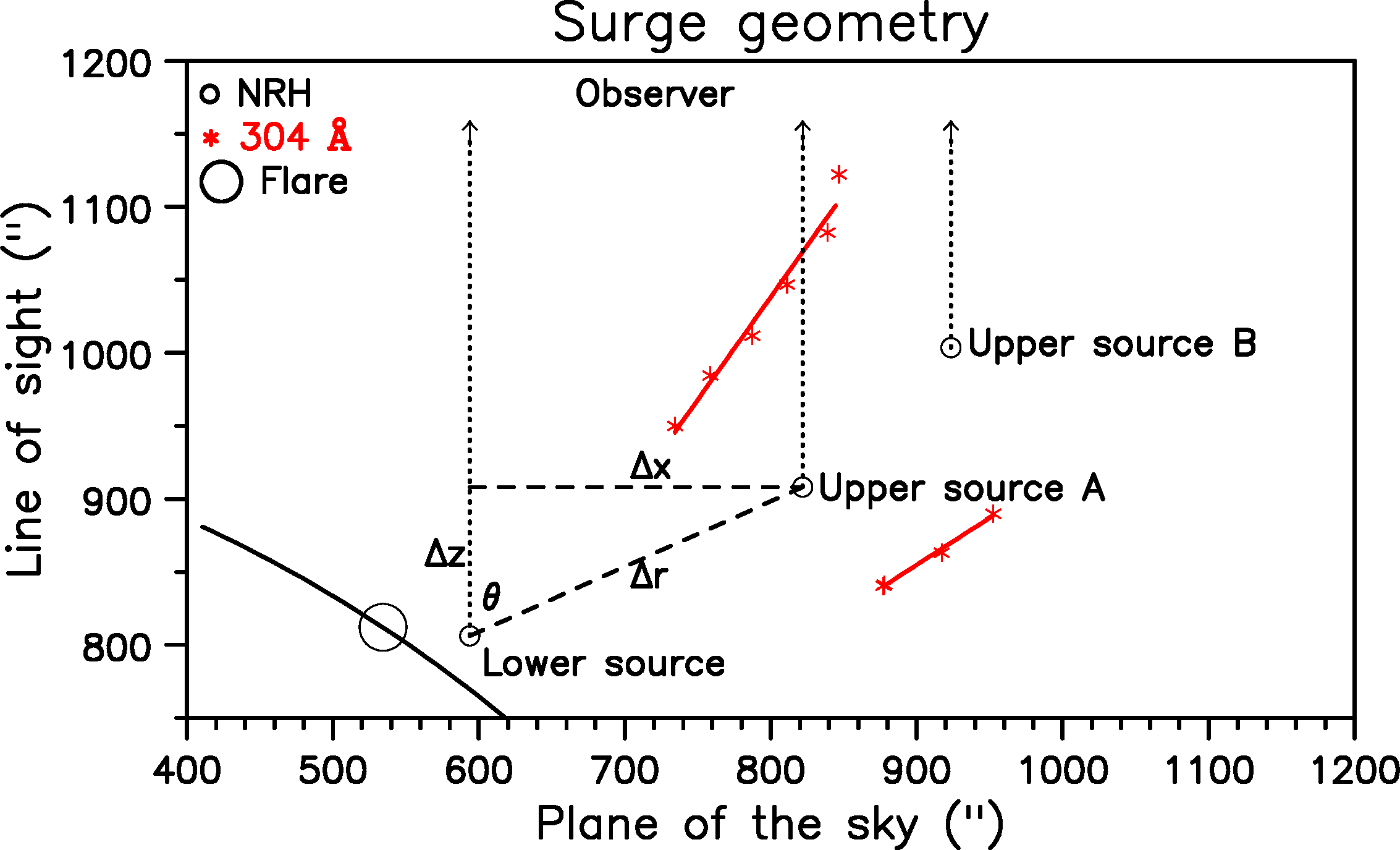}
\caption{NRH source positions on a plane perpendicular to the sky plane (open circles). A and B refer to the corresponding phases of the emission. The limits of the 304\,\AA surge derived from triangulation are also plotted in red for reference (c.f. Fig.~\ref{Inclin}). Dotted lines with arrows point towards the observer. The black arc marks the solar limb.}
\label{geometry}
\end{figure}

The apparent superluminal velocities deduced from the delay of fluctuations between the upper and lower sources during phase A (Sect.~\ref{PhaseA}) and the start of phase B (Sect.~\ref{PhaseB}) point towards the possibility that the emission of the upper source is scattered radiation from the lower source. In this section we will explore this possibility, with help of Fig.~\ref{geometry}.

In deducing the 3-dimensional positions of the NRH sources we assumed that the upper source was located in the middle of the region defined by the edges of the 304\,\AA\ surge, as derived from triangulation (red symbols and lines in Fig.~\ref{geometry}). For the lower source we assumed that it was located lower, at a height of 1.03\,R$_\sun$, consistent with the computations in the next section. Then the delay between emission from the lower source and scattered emission from the upper source is:
\be
\Delta t=\frac{\Delta r- \Delta z}{c}
\ee
and, taking into account the geometry.
\be
\Delta t=\frac{\Delta x}{c}\left(\frac{1}{\sin \theta}-\frac{1}{\tan \theta}\right)
\ee

For phase A the geometry gives $\theta\simeq24$\degr, whereas $\Delta x\simeq$ 230\arcsec = 165\,Mm, hence:
\be
\Delta t\simeq0.11\mbox{\,s}
\ee
which is compatible with the measured delay of 0.2\,s. 

Similarly, for the start of phase B we have $\theta\simeq31$\degr\ and $\Delta x\simeq$ 236\,Mm, thus:
\be
\Delta t\simeq0.21\mbox{\,s}
\ee
which is again compatible with the observed delay of 0.30\,s. We note that the assumed height of the lower source does not affect much the results; a greater height would increase $\theta$ slightly and decrease $\Delta t$.

We conclude that scattering of the emission from the lower source by the upper is a plausible interpretation of the apparent superluminal velocity. This conclusion is further strengthened by the decrease of the upper/lower intensity ratio with time, both between the two phases, as well as during phase B: it is a natural consequence of the increasing distance between the two sources, to which we may add the decrease of the density of the surge as it expands. The observed decrease of the upper/lower source intensity ratio at lower frequencies may imply that the scattering is more efficient at high frequencies
 
\subsection{Emission mechanism}\label{mech} 
Having established that the emission of the upper source was due to scattering from the lower source, which was located at or near the site of energy release, we will focus on the emission mechanism of the latter, in particular  to its high-frequency component ($\gtrapprox270$\,MHz) which was directly associated with the surge. Noting also that pulsations and spike-like features were present in this emission, we will consider only the component with slow time variations. In this section we will discuss all three possible emission mechanisms: thermal emission, gyrosynchrotron and plasma.

It is well known (\citeads[e.g. see the review by~]{2020FrASS...7...57N}) that in flare events the heated coronal plasma becomes optically thick to thermal free-free emission below about 10 GHz. In such case, the brightness temperature can reach values exceeding 10$^7$ K, its spectrum is flat and the degree of circular polarization is low ($\lessapprox10$\%). In our case the lower source exhibits flat spectra during phase A from 327 to 408 MHz (top panel of Fig.~\ref{SpecAB}) and during phase B in the interval 12:01-12:03 UT from 408 to 432 MHz (dashed line of bottom panel of Fig.~\ref{SpecAB}). However, the degree of circular polarization is higher than 80\% throughout the event which rules out the interpretation in terms of free-free emission. More important, there was not any surge-associated emission in the GOES data (Sect.~\ref{fluct}), whereas the SXR emission had dropped to preflare levels at the time of the surge (see Fig. 2 of Paper I), which precludes the presence of high temperature plasma; this also rules out the possibility of thermal gyroresonance emission.

We will next investigate whether gyrosynchrotron radiation from nonthermal electrons is at the heart of the emission. To this end we first attempted to model the high-frequency part of the spectrum from 12:01-12:03 UT in the early part of phase B (dashed line in the bottom panel of Fig.~\ref{SpecAB}) with a gyrosynchrotron spectrum. We used the inhomogeneous gyrosynchrotron code described in \citetads[][see also \citeads{2004A&A...420..351K}; \citeads{2008SoPh..253...79T}]{2000ApJ...533.1053N} with the addition that the effects of the ambient medium are taken into account, in order to properly account for the Razin-Tsytovich suppression (\citeads[e.g~]{razinv1960}; \citeads{1987A&A...183..341K}; \citeads{2008SoPh..253...43M}) at low frequencies.

We constrained our models by fixing both the source angular scale and line-of-sight length to 200\arcsec, in agreement with the results from the gaussian fit to the radio images (see Sect.~\ref{LowResAn}); these fits showed that the source size did not change appreciably with frequency, which justifies our decision  to work with the brightness temperature spectra instead of the flux density spectra. The remaining fit parameters are the energy spectral index, $\delta$, of the energetic electrons, the magnetic field, $B$, the low- and high-energy cutoffs, $E_1$ and $E_2$, respectively, of the nonthermal electrons, their density $N_r$, and the density of the ambient thermal plasma, $N_0$. The parameterized gyrosynchrotron model was fit to the spectrum using the Leveberg-Marquardt least-squares method. The best-fit model parameters, as well as their uncertainties, are given in Table~\ref{T01}.

\begin{table}
\begin{center}
\caption{Best-fit model gyrosynchrotron parameters}
\begin{tabular}{L{0.27\textwidth}c}
\hline 
Parameter                                                             & Value  \\
\hline
Energy spectral index of energetic electrons, $\delta$ & $2.7 \pm 0.1$ \\
Energy range of energetic electrons, $E_1-E_2$        & $(5 \pm 1)$-$(150 \pm 20)$\,keV            \\
Density of energetic electrons, $N_r$                       &  $(6.1 \pm 0.5) \times 10^6$\,cm$^{-3}$ \\
Density of ambient thermal plasma, $N_0$              &  $(1.5 \pm 0.8) \times 10^8$\,cm$^{-3}$ \\
Magnetic field strength                                           & $(5 \pm 2)$-$(35 \pm 5)$\,G                    \\
Characteristic angular scale of source                       & 200\arcsec\                                              \\
Line-of-sight length                                               & 200\arcsec\                                               \\
\hline 
\end{tabular}
\end{center}
\label{T01}
\end{table}

Our model gyrosynchrotron spectrum is shown as the blue dotted line in Fig.~\ref{SpecAB}. This model yields brightness temperatures which are broadly consistent with those observed from 270 to 445 MHz. However, the model degree of circular polarization ranges from 40\% to 70\% in the optically thin part of the spectrum whereas observations indicate that it is higher than 80\%. The discrepancy is even worse at 270.6 and 298.7 MHz, which fall into the optically thick part of the spectrum, where the model gives a degree of circular polarization as low as 25\%. It is possible to adjust the fit parameters to make the model polarization approach the observed one in the optically thin frequencies, at the cost of both unmatched total intensity brightness temperatures (differences of more than a factor of 4) and the shift of the peak emission to about 400 MHz.

The characteristics of the phase B spectrum from 12:05 to 12:10 UT (full line of the bottom panel of Fig.~\ref{SpecAB}),  in particular the higher brightness temperature and spectral peak around 360 MHz, are difficult to reproduce using any plausible combination of gyrosynchrotron model input parameters. The same is true for the spectrum of phase A (top panel of Fig.~\ref{SpecAB}), which shows an extended flat region at high frequencies that we could not reconcile with gyrosynchrotron models.

Using the gyrosynchrotron model parameters we can place constraints on the energetics of the surge. From the density of the ambient thermal plasma  (1.5$\times$10$^{8}$\,cm$^{-3}$) and the maximum height and speed of the surge of this study, $\approx$ 0.5\,R$_{\odot}$ and 130\,km\,s$^{-1}$, respectively, we found a surge mechanical (kinetic+potential) energy density of  $\approx0.02$\,erg\,cm$^{-3}$. The model magnetic field strength of the surge in the 5-35\,G range gives a magnetic energy density of $\approx$1-49\,erg\,cm$^{-3}$, which is more than sufficient to balance the mechanical energy density. This finding is consistent with a magnetic driving of the observed surge. We reached the same result by employing the density of the surge reported in \citepads{2013ApJ...770L...3K} (i.e., $4.1\times$10$^{9}$\,cm$^{-3}$), leading to a surge mechanical energy density of $\approx0.6$\,erg\,cm$^{-3}$, still sort of the magnetic energy density. We note that, in variance with \citepads{2013ApJ...770L...3K}, we were not able to derive the density from the Differential Emission Measure, since our data set lacked simultaneous multi-channel observations of the surge.

We now turn to the plasma mechanism. The observed high degree of circular polarization is consistent with plasma emission from the fundamental. It is well known that plasma emission is a leading candidate for type IV bursts (\citeads[see~][~and references therein]{2008SoPh..253....3N}). In several cases (see, e.g., \citeads{2015SoPh..290..219B}), these bursts exhibit fine structures that are similar to those detected in the dynamic spectrum of this surge (Figs.~\ref{DynSpecA} and \ref{DynSpecB}). We note that type IV emissions come with a variety of bandwidths, sometimes not extending below 300\,MHz (see events 17 and 20 in \citeads{2015SoPh..290..219B}), as is the case with the present event. Therefore, an interpretation in terms of type IV plasma emission could be consistent with these aspects of our event. 

\begin{figure}
\centering
\includegraphics[width=\hsize]{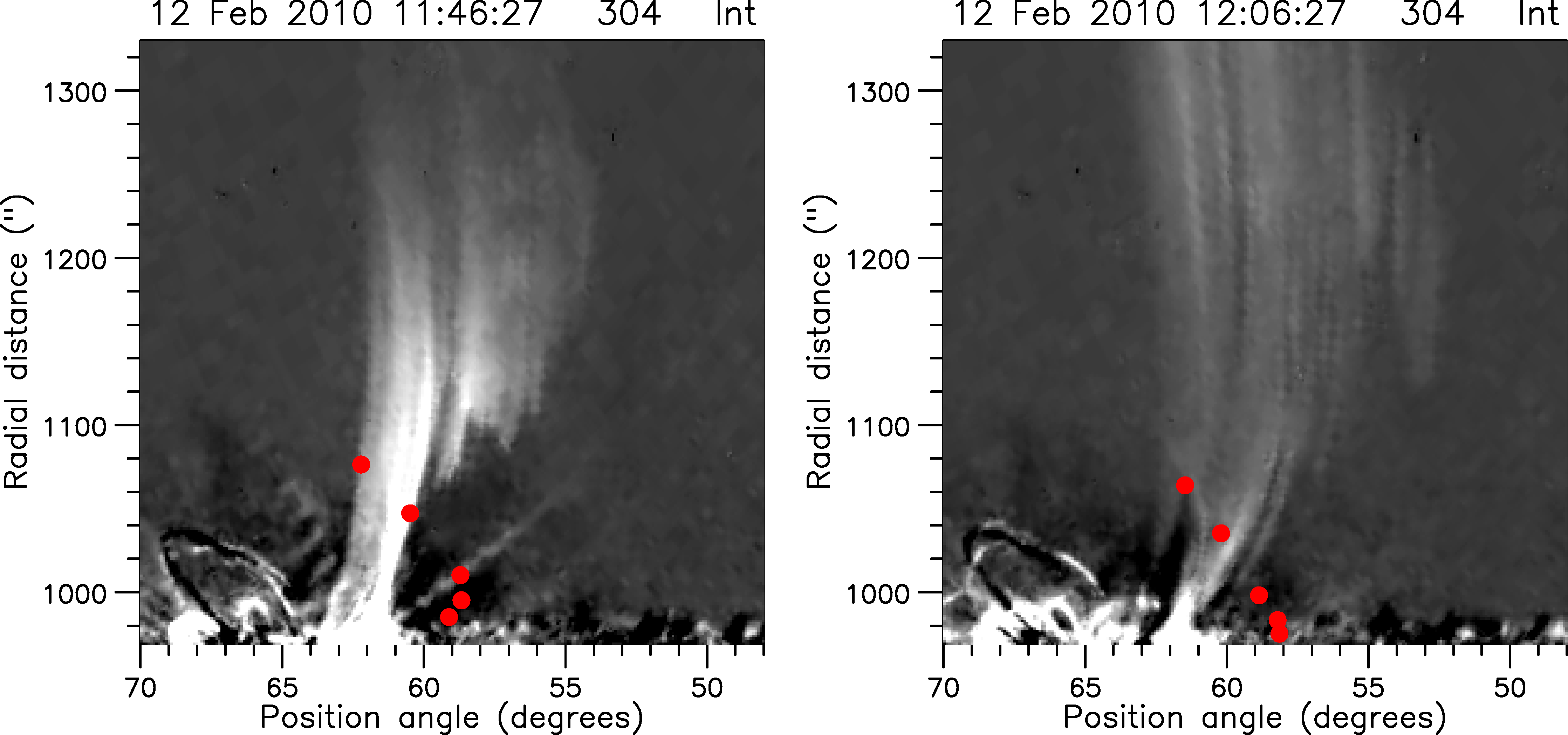}
\caption{Computed NRH source positions from 327 to 435\,MHz (red filled circles) projected on the sky plane as seen from STEREO-A, on top of base difference 304\,\AA\ images near phase A (left) and during phase B (right). Higher frequency sources are at lower radial distance.}
\label{304+NRH_Limb}
\end{figure}

Under the hypothesis of emission from the fundamental, the height of the radio sources can be computed. From that, their projected position above the limb as seen from STEREO-A  can be deduced and compared with the surge images at 304\,\AA. The results are displayed in Fig.~\ref{304+NRH_Limb}, for the two phases of the radio emission. The computation was performed for an isothermal coronal model at $2\times10^6$\,K and a base electron density 3.5 times that of the Newkirk model. We note that the projected positions are close to the base of the 304\,\AA\ surge, which adds strength to the type IV interpretation. { Our choice of the density multiplication factor, $k$ took into account the fact that for $k<3$ the 445\,MHz level is below the photosphere; going to $k=4$ gives a height difference of  15\,Mm, which would still put the NRH source near the base of the 304\,\AA\ surge.}

\section{Summary and conclusions}\label{concl}
In this work we reported the detection of metric radio emission from a surge, associated with a secondary energy release during the late phase of the M9 event of February 12, 2010, analyzed in Paper I. Although there have been previous reports of metric emission associated with SXR jets, this is the first time that direct imaging of a surge is reported.

The surge was imaged by STEREO A and B, seen near the east and west limbs respectively. Although no optical or EUV observations near the Earth were available and nothing was detected in SXR images, the identification of the NRH emission with the surge was established with confidence from triangulation of STEREO 304\,\AA\ and COR1 images. Moreover, consistent expansion velocities of the order of 150\,km\,s$^{-1}$ were measured from NRH and STEREO data. We note that no surge-associated emission had been detected during the main phase of that event, apparently due to its low brightness temperature and the limited dynamic range of the NRH.

The metric emission consisted of two spectral components, a high frequency component definitely associated with the surge and a low frequency one of different diverse, with a crossover around 270\,MHz. The high frequency component developed in two phases, A and B, phase B being much longer and brighter than phase A. Moreover, it consisted of two radio sources, one near the base of the surge (lower source) and one in the upper part (upper source), during both phases. The separation of these sources increased with time, reflecting the expansion of the surge, while the intensity of the upper source decreased, compared to that of the lower one.

Both sources showed short time scale variations, narrow-band and broad-band. There was a high degree of coherence between the upper and the low source during phase A and the beginning of phase B, with a delay of the upper source by 0.2 and 0.3\,s for A and B respectively. Although, at a first glance, this delay suggests superluminal velocities, a proper consideration of the geometry revealed that scattering of radiation from the lower source by the upper source is the most likely explanation. This scattering is more efficient at higher frequencies.

The two spectral components and the presence of short time scale fluctuations were confirmed in dynamic spectra, constructed from the 10 frequency channels of the NRH (``pseudo dynamic spectra'') as well as from the e-Callisto instrument of the Bleien Observatory and from ARTEMIS-JLS. They all show signs of pulsations and spike-like emission, in addition to the continuum components. 

Taking into account all available information, we examined possible emission mechanisms of the high-frequency component of the lower source which was apparently located at or near the site of energy release. 

The morphology of the total intensity spectra, the high degree of circular polarization and the absence of strong SXR emission rule out an interpretation in  terms of thermal mechanisms. 

The lower source total intensity spectrum during phase B from 12:01 to 12:03 UT was reproduced fairly well by a model of gyrosynchrotron emission from mildly relativistic electrons with energy cutoffs of 5 and 150\,keV and density of $6.1 \times 10^6$\,cm$^{-3}$ which, radiated in magnetic fields with strength ranging from 5 to 35\,G. However, this model yields degrees of polarization that are lower than the observed ones: 40-70\% versus more than 80\% in the optically thin part of the spectrum and even lower in the optically thick part. Attempts to fix this problem at high frequencies ruins the overall match between the observed and model spectrum whereas the degree of circular polarization mismatch would persist at low frequencies. At other time intervals the disagreement between observations and gyrosynchrotron models was even worse. 

We note that the parameters of the gyrosynchrotron model lead to the plausible conclusion that the energy stored in the magnetic field of the surge is sufficient to balance mechanical energy losses, a result pointing to a magnetic driver of the surge. We add that, even if gyrosynchrotron emission occurred for a short time interval only, the derived physical parameters and the energetics based on them would still reflect the actual situation.

It appears that the observed high degree of circular polarization, as well as the presence of spikes and pulsations in the dynamic spectrum, could be accommodated by plasma emission from the fundamental. Moreover, radio source positions computed under this mechanism are consistent with the surge position as seen from STEREO-A. We hence consider type IV-like plasma emission with a low intensity gyrosynchrotron component as the most plausible mechanism.

The detection of surge-associated emission adds one more manifestation to the list of flare-associated metric radio emissions. More observations of this kind will help establish its characteristics and the nature of the emission.

\begin{acknowledgements} 
The authors wish to thank the radio monitoring service at LESIA (Observatoire de Paris) for providing data used for this study, in particular our colleagues from the Observatoire de Paris-Meudon and the NRH staff. Data from CALLISTO, GOES, NOAA and STEREO were obtained from the respective data bases; we are grateful to all those who contributed to the operation of these instruments and made the data available to the community.
\end{acknowledgements}

% -----------------------------------------------------------------

\begin{appendix}

\begin{figure*}[]
\centering
\includegraphics[width=\textwidth]{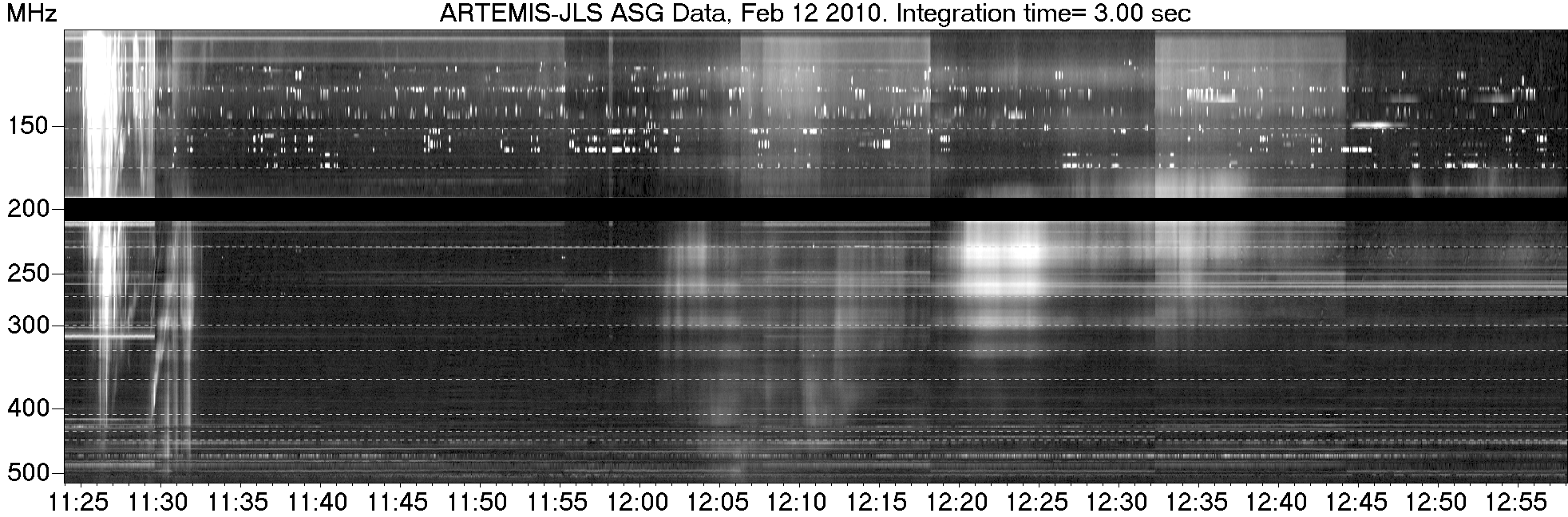}
\caption{{ Low time resolution dynamic spectrum of the entire event in the long decimetric-metric range from 11:24 to 12:58 UT. Dashed horizontal lines mark the NRH frequencies.}}
\label{FullDS}
\end{figure*}

\begin{figure*}[]
\centering
\includegraphics[width=\textwidth]{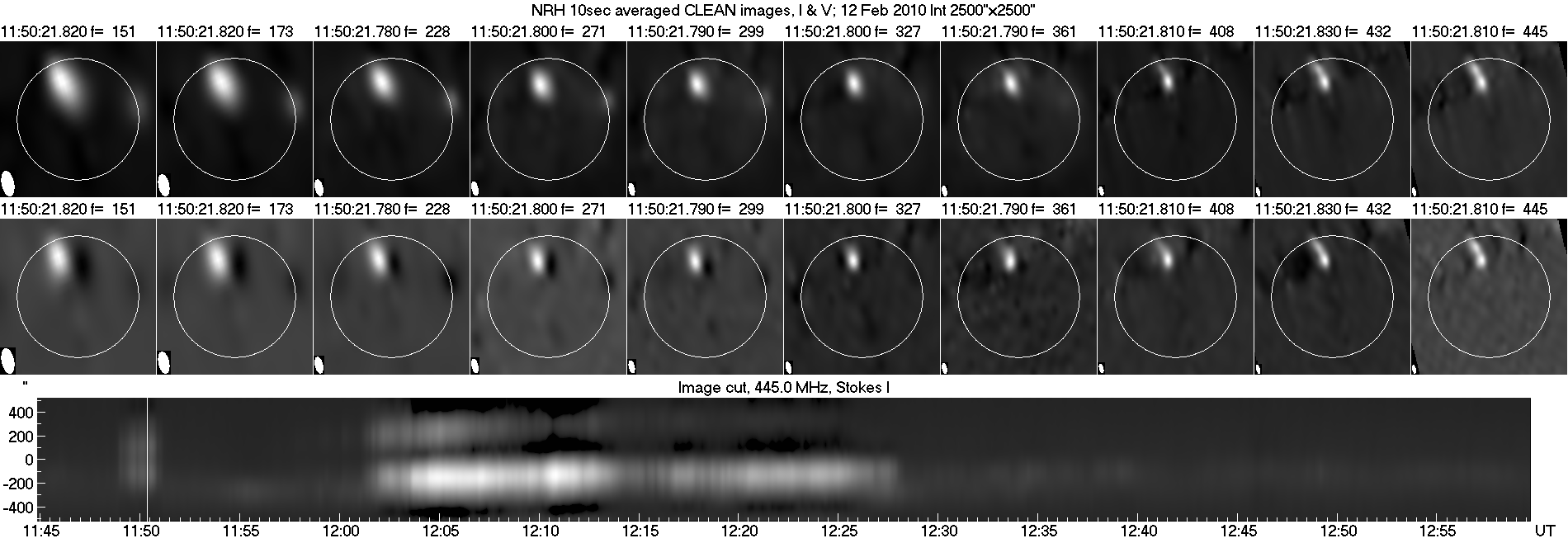}
\vspace{.4cm}
\includegraphics[width=\textwidth]{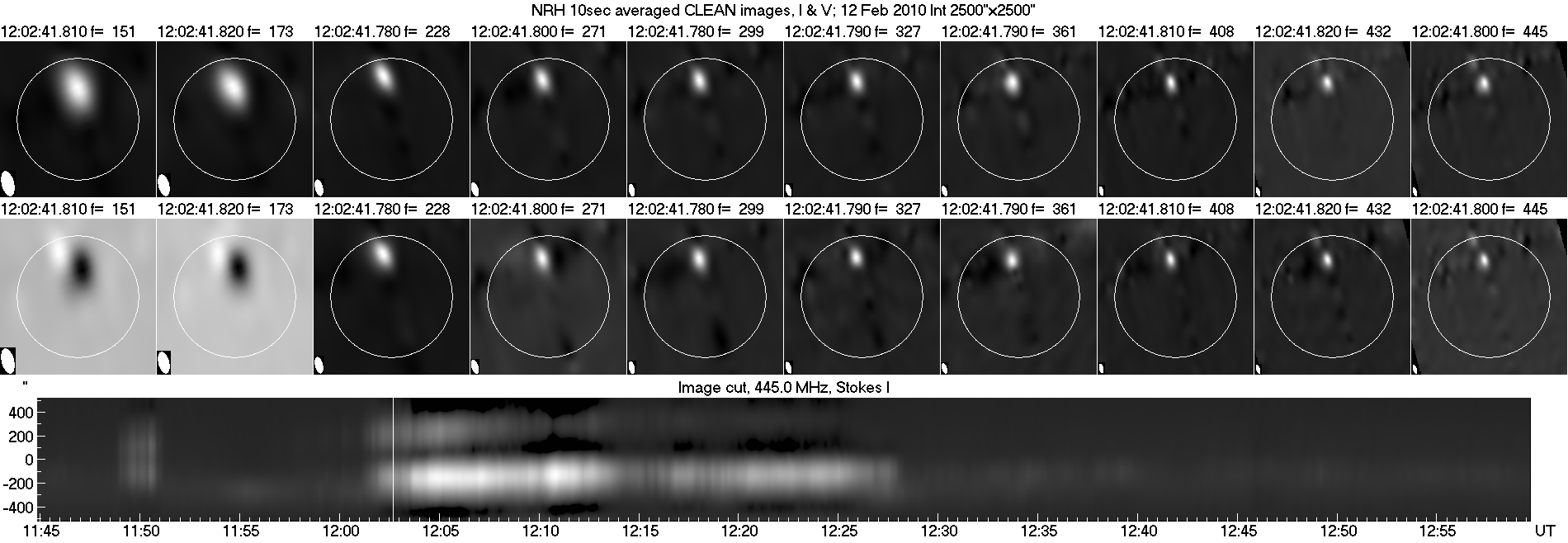}
\caption{Two frames of Movie 1, during phase A and at the beginning of phase B. Each frame shows radio images at all 10 NRH frequencies in total intensity (top row) and circular polarization (middle row). The color table of each image is such that the maximum intensity is white and the minimum intensity is black. The instrumental beam is shown in the low left corner, while the white circle marks the photospheric limb. In the bottom row a cut of the intensity along the surge as a function of position and time is given, where the white vertical line marks the time of the images.}
\label{frames}
\end{figure*}

\section{Supplementry material}\label{Appendix}
\subsection{{ Full dynamic spectrum}}
{ Fig.~\ref{FullDS} shows the entire dynamic spectrum of the February 12, 2010 event (c.f. Fig. 45 of \citeads{2015SoPh..290..219B}). The emission before 11:33 UT is from the main event described in paper I (c.f. Fig. 1 in \citeads{2021A&A...654A.112A}), with the 2 type IIs and numerous type IIIs. The emission after 12:00 UT is from phase B of the surge, whereas emission from phase A was below the threshold of the instrument.}

\subsection{Movie with 10\,s average NRH images}
Fig.~\ref{frames} shows two frames from Movie 1, which includes all 10\,s averaged NRH images in the interval from 11:45 to 13:00 UT, for both total intensity (Stokes parameter $I$) and circular polarization (Stokes parameter $V$). The intensity as a function of time and position along the surge at 445\,MHz is shown in the bottom frame, with the vertical line marking the time of the images.  
 
\end{appendix}

\end{document}